\documentclass[preprint,3p,times]{elsarticle}
 \usepackage{color}
  \usepackage[colorlinks=true]{hyperref}

\usepackage{amssymb}
\usepackage{amsthm}

\usepackage{graphicx}
\usepackage{float}
\restylefloat{figure}
\usepackage{epsfig}
\usepackage{amsmath}
\usepackage{natbib}
\usepackage{txfonts}
\usepackage{ifvtex}
\usepackage{amssymb}
\usepackage{amsfonts}
\usepackage{xfrac}
\usepackage{booktabs}
\usepackage{color}
\usepackage[subrefformat=parens, caption=false]{subfig}
\biboptions{numbers,sort&compress}

\begin{document}
	
	\begin{frontmatter}
		
		
		
	\title{ A class of charged-Taub-NUT-scalar  metrics via Harison and Ehlers Transformations}
		
		
		\author{Mahnaz Tavakoli Kachi}%
	\ead{m.tavakoli1399phy@gmail.com}
	\author{Behrouz Mirza}
	\ead{b.mirza@iut.ac.ir}
	\author{Fatemeh Sadeghi}%
	\ead{fatemeh.sadeghi96@ph.iut.ac.ir}

	\address{Department of Physics, Isfahan University of Technology, Isfahan 84156-83111, Iran}
	\date{\today}

		
		\begin{abstract}
		We consider a class of axially symmetric solutions to Einstein's equations incorporating a $\theta$-dependent scalar field and extend these solutions by introducing electric and magnetic charges via Harrison transformations. Subsequently, we enhance the charged metrics by incorporating the NUT parameter through Ehlers transformations, yielding a novel class of charged-Taub-NUT  metrics that represent exact solutions to Einstein's equations. Finally,  we investigate some of astrophysical aspects  of the charged-Taub-NUT  metrics, focusing on phenomena such as gravitational lensing and quasi-normal modes (QNMs).
			
		\end{abstract}
		


		
	\end{frontmatter}
	
	
	\section{\label{Introduction}Introduction}
The gravitational collapse of stars can lead to different final states, most notably the formation of black holes \cite{ hawking1972black, novikov1973astrophysics, carr1974black, preskill1992black, taylor2000exploring, alexander2012drives, novikov2013physics} or naked singularities \cite{penrose1973naked, shapiro1991formation, joshi1992structure, christodoulou1994examples, de2001turning, harada2002physical, goswami2006quantum, joshi2009naked}. 
	 The characteristics of these final states, such as rotation parameters and electric charge, depend on the initial conditions of the collapsing star.
	
	 But how can charged final states form?  Given the extreme gravitational fields, pressures, and densities within stars, charge distribution directly related to matter density can give rise to strong electric fields and substantial charges. The forces acting on charged particles within the star may eject these particles, disrupting equilibrium. Once the gravitational force surpasses the pressure of matter and the Coulomb force, collapse occurs, and a charged singularity is formed. For the resulting electric charge and field to visibly impact the surrounding environment, their magnitudes must be significant  
	\cite{hawking1971gravitationally, sorkin2001formation, ray2003electrically, cuesta2003charge, hwang2011internal, hwang2012dynamical}. 
	The simplest example of a charged black hole is the Reissner-Nordström (RN) black hole, characterized by a static and spherically symmetric spacetime.
	
	A black hole can also have a gravomagnetic monopole, such a metric was first introduced in 1963 by Newman, Unti and Tamburino (NUT)
	\cite{newman1963empty, kramer1983exact}.  The published NUT metric  preceded the discovery of the Kerr metric and also paved the way for   its discovery \cite{kerr1963gravitational,newman2014kerr}.   A solution to Einstein's vacuum equations, the NUT metric remains a compelling subject of study without requiring modifications to Einstein's action   \cite{ramaswamy1986comment}.  The NUT solution  has found applications in higher-dimensional semi-classical theories of quantum gravity. Misner was among the first to provide a physical interpretation of the NUT metric, suggesting periodic time coordinates and closed timelike world lines for rest-frame observers 
	\cite{misner1963flatter, misner1967contribution}. 
	He intended to remove the singular regions from the metric and for this he considered periodic time coordinates and stated that every observer in the rest frame moves on the closed time like world line.  
	In 1969, Bonnor proposed a simpler interpretation, describing the NUT metric as a symmetric spherical mass connected to a semi-infinite source of angular momentum along its symmetry axis
	\cite{bonnor1969new}. 
	This interpretation portrays the NUT metric as a finite static rod with positive mass, connected to counter-rotating semi-infinite sources with infinite angular momenta and negative masses
	\cite{manko2005physical}.

A particularly intriguing aspect of the Taub-NUT spacetime is its geodesic similarity to the trajectories of charged particles in a magnetic monopole field.  \cite{zimmerman1989geodesics}.    Light rays passing near the origin not only bend but also twist due to the gravomagnetic force, making gravitational lensing a promising tool for detecting gravomagnetic masses   \cite{lynden1998classical,nouri1997gravomagnetic}.  The absence of direct evidence for gravitomagnetic masses may stem from the limited sensitivity of current instruments \cite{rahvar2003gravitational}. However, advancements such as the proposed X-ray observational technique could pave the way for their detection, opening new frontiers in physics  \cite{chakraborty2018does}. The Taub-NUT metric has also been explored in Euclidean gravity, where it could provide insights into quantum gravity  \cite{hawking1977gravitational,arratia2021hairy,barrientos2022gravitational}. Brill was the first to derive the charged-Taub-NUT metrics, which remain exact solutions to Einstein's equations and hold significant physical interest    \cite{brill1964electromagnetic}.  
	
	Ernst demonstrated that the Einstein-Maxwell theory could be reformulated for stationary, axially symmetric spacetimes using two gravitational complex potentials, leading to transformations like those of Ehlers and Harrison \cite{ernst1968new, ernst1968new2}.
	Ehlers transformations are vacuum symmetries. However one may use 	 the Ehlers transformations for   Einstein’s equations coupled with a scalar field 	\cite{astorino2013embedding,astorino2015stationary}.
 In this case,  the field equations reduce to two Ernst's complex equations and  an independent  equation for a massless scalar field \cite{astorino2013embedding}.   These transformations have been used to generate novel solutions, including accelerating NUT black holes coupled to a scalar field \cite{barrientos2023ehlers} and rotating wormholes in the Barceló-Visser spacetime \cite{cisterna2023exact}.   Moreover,   the Ehlers transformation was applied to Kerr-Newman black holes. In this case since the seed metric contains an electromagnetic field, the vector potential also  rotates due to Ehlers transformation \cite{astorino2020enhanced}.  
	
	 The Zipoy-Voorhees  (ZV) metrics are a class of exact solutions of Einstein’s equations with naked singularities \cite{darmois1927memorial, erez1959gravitational, zipoy1966topology, voorhees1970static}.  Observations of the Sagittarius A* shadow are consistent and  can be explained  with assuming ZV metrics   \cite{lora2023q}.     Gravitational waves and geodesics  of the class of ZV metrics have  been analyzed in  \cite{destounis2023geodesics}. The following articles    investigate some other aspects of ZV metrics \cite{richterek2002einstein,chakrabarty2018unattainable,toshmatov2019harmonic,allahyari2019quasinormal,chakrabarty2022effects}. Another class of exact  solutions of Einstein’s equations in the presence of a scalar field is Fisher-Janis-Newman-Winicour (FJNW) solutions 
	 \cite{fisher1999scalar, janis1968reality, wyman1981static}. Astrophysical properties such as gravitational lensing and shadows of FJNW metric was investigated in \cite{virbhadra2002gravitational,chen2024gravitational,solanki2022shadows}. Quasi-normal modes of FJNW metric in the presence of a non-linear scalar field was obtained in \cite{stashko2024quasinormal}. Other properties of FJNW metric including nature of singularity \cite{virbhadra1997nature} and   geodesics
	 \cite{ chowdhury2011circular} have been also studied.  The study of FJNW metrics has similarly advanced our understanding of singularity formation.  A recent work has introduced a three-parameter family of metrics   \cite{azizallahi2024three},  ZV
	and FJNW solutions. 
	Also, the rotating form of these metrics derived in 
	\cite{mirza2023class}, 
	that in certain limits of parameters represent rotating  ZV metric
	\cite{toktarbay2014stationary, frutos2018relativistic}, 
	Bogush-Gal’tsov (BG) metric
	\cite{bogush2020generation}, 
	and rotating FJNW metric.   
	 
	Motivated by these findings, we investigate a new two-parameter class of static metrics involving a $ \theta $-dependent scalar field with an event horizon   \cite{mazharimousavi2023nonspherically}. Using Harrison \cite{harrison1968new} and Ehlers \cite{ehlers1958konstruktionen} transformations, we extend these metrics by introducing charge and the NUT parameter, respectively.
	
	This work examines the astrophysical properties of charged-Taub-NUT metrics.  Specifically, we explore  light deflection due to gravitational lensing. There are various techniques to investigate gravitational lensing
	\cite{schneider1992gravitational, wambsganss1998gravitational, narayan1999gravitational, schneider2006gravitational, straumann2012general}. 
	 Another method to investigate the characteristics of the class of metrics is to study their corresponding QNMs. QNMs are related to waves that are propagated by disturbances created in the metric of black holes. These waves change over time and are damped
	\cite{nollert1996significance, nollert1999quasinormal, kokkotas1999quasi, dreyer2003quasinormal, berti2009quasinormal, denef2010black, konoplya2011quasinormal, flachi2013quasinormal, matyjasek2017quasinormal}. 
	Here, we use the light ring method in the eikonal limit to study QNMs
	\cite{ferrari1984new, ferrari1984oscillations, mashhoon1985stability}. By examining these phenomena, we aim to uncover deeper insights into the nature of the class of exact solutions.
	
	This paper is organized as follows: In Sec. \ref{charge}, we consider a class of  two parameters static metrics in the presence of  $ \theta $-dependent scalar field. Then,  we add electric and magnetic charges to the  metrics using Harrison's transformations. In Sec. \ref{4}, we apply Ehlers transformations to incorporate the NUT parameter, obtaining new exact solutions of Einstein's field equations.  Sec. \ref{5} is dedicated to analyzing gravitational lensing effects in the charged-Taub-NUT metrics. In Sec. \ref{6}, we investigate the QNMs of the metrics, using the eikonal limit.  Finally, we summarize our findings in  Sec. \ref{7}.
	\section{A  charged metric in the presence of a $ \theta $-dependent scalar field}\label{charge}
	There have been an increasing  motivation   to   obtain exact solutions of the  Einstein's  equations in the presence of both electromagnetic and  scalar fields. The Einstein-Hilbert action $ ( G = c = 1) $ with
	an electromagnetic field and a massless scalar field $ \psi $ is in the following form
	\begin{align}  
		S=\int {{d^4}} x\sqrt { - g} \, (R - {F_{\mu \nu }}{F^{\mu \nu }} - {g^{\mu \nu }}{\partial _\mu }\psi \,{\partial _\nu }\psi ).\label{eq:act}
	\end{align}
	where, $ F_{\mu \nu }=\partial_{\mu} A_{\nu}-\partial_{\nu} A_{\mu}$.   Varying the  action, the  field equations are given by
	\begin{align}  
		R_{\mu \nu}={\partial _\mu }\psi \,{\partial _\nu }\psi+T_{\mu \nu }^{EM},	\,\,\,\,\,\,\,\,\,\,\,\,\,\, \square	\psi=0,\label{eq:mov}
	\end{align}
	where, $ T_{\mu \nu }^{EM} $ defined as follows
	\begin{align}  
		T_{\mu \nu }^{EM}=2 \, {F_\mu }^\alpha {F_{\nu \alpha }} - \frac{1}{2} \, {g_{\mu \nu }}{F_{\gamma \tau }}{F^{\gamma \tau }},
	\end{align}
	In the case that  the electromagnetic energy-momentum tensor is zero i.e. $ 	T_{\mu \nu }^{EM}=0  $,  two classes of  exact solutions can be obtained  from  Eq. \eqref{eq:mov} depending on  whether  a massless scalar field $ \psi $ is a function  of  $ r  $  or $ \theta  $. The general class of  exact solutions  in the presence of  a massless scalar field  $ \psi (r) $ that recently found in \cite{azizallahi2024three} is in the following form  
	\begin{align}  
		ds^{2}=-f^{\gamma} dt^{2} + f^{\mu}  k^{\nu} (\frac{dr^{2}}{f}+r^{2} d\theta^{2})+ f^{1-\gamma} r^{2} \sin^{2} \theta \, d\phi^{2},\label{eq:metric3}
	\end{align}
	where,
	\begin{align} 
		f(r)=1-\frac{2m}{r}, \,\,\,\,\,\,\,\,\,
		k(r,\theta)=1-\frac{2m}{r}+\frac{m^{2} \sin \theta^{2}}{r^{2}},
	\end{align}
	and  
	\begin{align} 
		\mu+\nu&=1-\gamma.
	\end{align}
	with the following scalar field
	\begin{align}  
		\psi(r)=\sqrt{\frac{1-\gamma^{2}-\nu}{2}} \ln (1-\frac{2 m}{r}).
	\end{align}
	The three parameter metrics in Eq. \eqref{eq:metric3}  for $ \nu =0$ and $ \mu=1-\gamma $  represent the FJNW metric.  Moreover, Eq. \eqref{eq:metric3} by choosing $ \mu=\gamma^{2}-\gamma $ and $ \nu=1-\gamma^{2} $  represents the ZV metric. It is interesting that the static solution in Eq. \eqref{eq:metric3} can be obtained by applying  Buchdahl's transformation of the second kind to the metric of  ZV spacetime \cite{barrientos2024revisiting}. 
	The second class of  solutions that satisfy  the
	equations of motion with a  massless   $  \theta$-dependent scalar field $ \psi (\theta) $ is in the following form \cite{mazharimousavi2023nonspherically}
	\begin{equation}
	ds^{2}=-f(r) \, dt^{2}+ k(r,\theta)^{\sigma} (\frac{dr^{2}}{f(r)}+ r^{2} d\theta^{2})+ r^{2} \sin ^2\theta \, d\phi^{2},\label{eq:metric}
	\end{equation}
	where,
	\begin{equation}
	f(r)=1-\frac{2 \, m}{r},\;\,\,\,\,\,\,\, k(r,\theta)=\frac{m^2 \sin ^2\theta }{r^2 -2 \, m \, r+m^2 \sin ^2\theta},\label{eq:f1}
	\end{equation}
	where, the corresponding scalar field  is in the following form
	\begin{equation}
	\psi(\theta)=\sqrt{2 \, \sigma } \ln \left(\tan \frac{\theta }{2}\right),
	\end{equation}
	In the following sections,  we investigate some properties of metric in  Eq. \eqref{eq:metric}.  The class of  metrics in Eq. \eqref{eq:metric} have   an axial singularity at two points $ \theta =0 $ and $ \theta =\pi $. However an event horizon exists for  other values of $\theta$ parameter.  The singular points coincides with those of the Taub-NUT black holes.
	 To derive the charged form of the metric, we use Harison transformation. We review   the algorithm for obtaining the Ernst equations and Harison transformation. We will show that the field equations related to the scalar field decouples from  the other equations and therefore  in the presence of a scalar field   the  Harison and Ehlers transformations can be used in their original form, to generate new solutions, see \cite{astorino2013embedding} for further details. Consider the   Lewis-Weyl-Papapetrou metric (LWP metric) in the following form
	\begin{equation}
	ds^{2}=-f( dt-\omega \,  d\phi)^{2}+ f^{-1} (e^{2 \, \eta}(d\rho^{2}+ dz^{2})+\rho^{2} d\phi^{2}),	\label{eq:LWP}
	\end{equation}
	where, the $ \rho $  and $ z $ are   functions of $ r $  and $ \theta $ as follows
	\begin{equation}
	\rho=\sqrt{ r^{2} -2 \, m \, r } \, \sin \theta,\,\,\,\,\,\,   z=(r-m) \, \cos \theta.\label{eq:ct}
	\end{equation}
	The compatible vector potential according to the symmetry of  LWP metric  is defined as below
	\begin{equation}
	A= A_{t} \, dt + A_{\phi}\,  d\phi.\label{eq:vector}
	\end{equation}
	By considering action in    Eq. \eqref{eq:act}  and  metric in Eq. \eqref{eq:LWP},  the gravitational equations of motion can be written as follows 
	\begin{align}  
		f \, \nabla^{2} f= (\nabla f)^{2}-\rho^{-2} f ^{4} (\nabla \omega)^{2}+ 2 f (\nabla A_{t})^{2} + 2 \,\rho^{-2} f ^{3} (\nabla A_{\phi}+ \omega \nabla A_{t})^{2}, \label{eq:max1}
	\end{align}
	\begin{align}  
		\nabla. [\rho ^{-2 } f^{2} \,  \nabla \omega + 4 \, \rho^{-2} \, f \, A_{t} (\nabla A_{\phi}+ \omega \nabla A_{t})]=0 , \label{eq:max2}
	\end{align}
	whereas the Maxwell and scalar equations are  as follows
	\begin{align}  
		\nabla.[-f^{-1}  \nabla  {A_{t}} + \rho ^{-2 } \, f \omega \,  (\nabla A_{\phi}+ \omega \nabla A_{t})] =0,\label{eq:eq1}
	\end{align} 
	\begin{align}  
		\nabla.[ \rho^{-2}  \, f (\nabla {A_{\phi}}+ \omega \, \nabla A_{t} )]=0, \label{eq:eq2}
	\end{align} 
	\begin{align}
		\square \psi=0.\label{eq:phi1}
	\end{align} 
	It is seen  that the scalar field equation is an independent equation  which is not coupled to the gravitational and the Maxwell field equations so Ernst's equations remain the same,  more explanation can be found in \cite{astorino2013embedding}.  
	Ernst proposed that  the Einstein-Maxwell field equations i.e.  Eqs. \eqref{eq:max1}-\eqref{eq:eq2}   can be written    in the form of two simple complex equations \cite{ernst1968new, ernst1968new2}.  Eqs \eqref{eq:max1}-\eqref{eq:phi1}  can be written  as the following two Ernst's complex equations in addition to a scalar field equation as follows
	\begin{equation}\label{h4_label}
	\begin{aligned}
	\big(Re(\mathcal{E})+\Phi\,\Phi^\star\big)\,\nabla^2\,\mathcal{E} &= \nabla\,\mathcal{E}\, .\, (\nabla\,\mathcal{E}+2\,\Phi^\star\,\nabla\,\Phi),\\
	\big(Re(\mathcal{E})+\Phi\,\Phi^\star\big)\,\nabla^2\,\Phi &= \nabla\,\Phi\, .\, (\nabla\,\mathcal{E}+2\,\Phi^\star\,\nabla\,\Phi),  \\ \square \psi&=0,\\
	\end{aligned}
	\end{equation}
	where, $ Re(\mathcal{E}) $ is the real part of the complex gravitational potential and the gravitational potential $ \mathcal{E} $ can be defined as follows
	\begin{align}
		\mathcal{E}:=f-\left| \Phi  \right|^{2} + i \, \chi,\label{eq:varepsilon}
	\end{align}
	where, $ f $ is the metric coefficient. Also, other functions of $ \Phi$ and $ \chi$  are written as
	\begin{equation}\label{eq:phichi}
	\begin{aligned}
	&\Phi=A_t+i\,\tilde{A}_\phi,\\
	&\hat{\phi}\times\nabla\,\chi=-\rho^{-1}\,f^2\,\nabla\,\omega-2\,\hat{\phi}\times Im(\Phi^\star\,\nabla\,\Phi),\\
	\end{aligned}
	\end{equation}
	where, $ Im(\Phi^\star\,\nabla\,\Phi) $ is the imaginary part of the  $ \Phi^\star\,\nabla\,\Phi $ and $  \tilde A_{\phi} $ is a twisted potential.   The components $ A_{t} $ and $  A_{\phi} $  are   the first and  fourth component of the 4-vector potential respectively, and  satisfy the following relation   
	\begin{equation}\label{eq:AtAphi}
	\hat{\phi}\times\nabla\,\tilde{A}_\phi=\rho^{-1}\,f\,(\nabla\,A_\phi+\omega\,\nabla\,A_t),
	\end{equation}	
	Now, by using the  coordinate transformation in Eq. \eqref{eq:ct},   we find $ f, \omega  $ and $ e ^{2\eta} $  as follows
	\begin{align}  
		f(r)= (1-\frac{2 \, m}{r}), \,\,\,\,\,\,
		\omega=0,\label{eq:LWP1} 
	\end{align} 
	\begin{align}
		e ^{2 \, \eta}&=\frac{{ (r^{2}-2\, m \, r)}  \, m^{2 \, \sigma} \, \sin^{2 \, \sigma}\theta} {({r^{2}-2\, m \, r+m^2 \sin^2\theta}) \, ^{\sigma+1}} .\label{eq:LWP2}
	\end{align}
	The Harrison transformations are defined in the following form
	\begin{align}
		\mathcal{E}   ' &=\frac{\mathcal{E}}{1-2 \,\alpha^{\star} \, \Phi - \left| \alpha ^2\right| \, \mathcal{E} }, \nonumber\\	
		\Phi  ' &=\frac{\alpha \, \mathcal{E}+  \Phi }{1-2 \, \alpha^{\star} \, \Phi - \left| \alpha ^2\right| \, \mathcal{E}},\label{eq:HT}
	\end{align} 
	where, $ \alpha $ is a complex parameter.  According to the metric in Eq. \eqref{eq:metric}  and using Eq. \eqref{eq:varepsilon}, Eq. \eqref{eq:phichi} and Eq. \eqref{eq:LWP1}     we choose the following functions
	\begin{align}
		\mathcal{E}   =1-\frac{ 2 \, m}{r},\,\,\,\,\,\,\,
		\Phi   =0, \,\,\,\,\,\,\, \chi=0.\label{eq:E1}
	\end{align}
	Then, by inserting Eq. \eqref{eq:E1}  in Eq. \eqref{eq:HT}  we have
	\begin{align}
		\mathcal{E}   ' &=\frac{(r-2 \, m)}{r-\left| \alpha ^2\right| (r-2 \, m)}, \nonumber\\	
		\Phi  ' &=\frac{ \alpha \, (r-2 \, m) }{r-\left| \alpha ^2\right| (r-2 \, m)}.\label{eq:ph}
	\end{align}		
	Therefore,  using Eq. \eqref{eq:varepsilon}  and Eq. \eqref{eq:ph}  we obtain $ f'$ in the following form
	\begin{equation}
	f  ' = \frac{r (r-2 \, m)}{(r-(r-2 \, m) \left| \alpha ^2\right| )^{2}},\,\,\,\,\,\, \,\,\,
	\chi '=0.\label{eq:fprim}
	\end{equation}
	Under Harrison transformation the parameter $\eta  $  in Eq. \eqref{eq:LWP2} remain the same so  $ \eta=\eta' $. By considering  $ \alpha $ as  a complex parameter $ \alpha =\alpha_{R} + i \alpha_{I} $ and  Eq. \eqref{eq:ph}  and  Eq. \eqref{eq:vector} we arrive at 
	\begin{equation}
	{A'_t}= \frac{\alpha_{R} \, (r-2 \, m)}{r-\left| \alpha ^2\right|\, (r-2 \, m) },\,\,\,\,\,\,	{\tilde A_\phi '}= \frac{\alpha_{I} \, (r-2 \, m)}{r-\left| \alpha ^2\right|\,(r-2 \, m)  },\label{eq:Apm}
	\end{equation}
	where, $ {\tilde A_\phi '} $ is a twisted potential. Now,  to determine ${ A'_\phi } $   we use   Eq. \eqref{eq:AtAphi}.  Since  we  are in  the  cylindrical  coordinates $ (\rho, \phi, z) $,  to use   the spherical coordinate,      we  use   the following  transformation  relation for gradient 
	\begin{equation}
	\nabla F (r,\theta)= \frac{1}{\sqrt{(r-m)^2-m^2 \cos^2\theta }}(\frac{{\partial \,F (r,\theta)}}{{\partial \,r}} \sqrt{r (r-2 \, m)}\,\hat r + \frac{{\partial \,F (r,\theta)}}{{\partial \,\theta }}\,\hat \theta ),\label{eq:grad}
	\end{equation}
	therefore, by using Eqs.    \eqref{eq:fprim}, \eqref{eq:Apm}, and Eq. \eqref{eq:grad} in  Eq. \eqref{eq:AtAphi} we have 
	\begin{equation}
	{ A_\phi '}= 2 \, m \, \alpha_{I} \, \cos \theta. \label{eq:Aprim}
	\end{equation}
	So the vector potential takes the fallowing form
	\begin{equation}
	{A'(r, \theta) }=\frac{\alpha_{R}  \, (r-2 \, m)}{r-\left| \alpha ^2\right| \, (r-2 \, m)  }\, dt+ 2 \, m \, \alpha_{I} \, \cos \theta \, d\phi. \label{eq:Aprimf}
	\end{equation}
	By using Eqs. \eqref{eq:LWP},  \eqref{eq:ct}  \eqref{eq:LWP2} and Eq. \eqref{eq:fprim}   the new form of metric  in Eq. \eqref{eq:metric} in  the presence of the electromagnetic field can be written as follows
	\begin{align}\label{eq:barmetric}
		{ds'^{2}}&=-\frac{r \, (r-2\, m)}{\left(r-\left|\alpha ^2 \right| (r-2 \, m)\right)^2}dt^{2} + \frac{\left(r-\left|\alpha ^2 \right| (r-2 \, m)\right)^2}{r \, (r-2 \, m)}\left(\frac{m^2 \sin ^2\theta }{ {r^{2}-2 \, m} \, {r}+{m^2 \sin ^2\theta }}\right)^{\sigma } \\ & \times (dr^{2} + r \,\, (r-2 \, m) \, d \theta^{2})\nonumber+ \left(r-\left|\alpha ^2 \right| (r-2 \, m)\right)^2 \sin ^2\theta \, d\phi^{2}. 
	\end{align}
	Now, we apply the following transformations on   coordinates and parameters in Eq. \eqref{eq:barmetric}  
	\begin{align}\label{eq:ctp1}
		\bar r &= r \, (1- \left|\alpha ^2 \right| ) + 2 \, m \left|\alpha ^2 \right|, \,\,\,\, \,\,\,\, \bar t=\frac{t}{1- \left|\alpha ^2 \right| },
		\nonumber\\ 	\bar m &= m \, (1+  \left|\alpha ^2 \right| ), \,\,\,\,\,\,\,\, q_{e}=-2 \, m \, \alpha_{R},  \,\,\,\,\,\, \,\,   q_{m}=2 \, m \, \alpha_{I},
	\end{align}
	where $ q_{e}  $ and  $ q_{m}  $ are electric and magnetic charges, respectively. Therefore, we arrive at the following charged metric, where  for simplicity  the parameters are rewritten as  ($ \bar r \to r, \,\, \bar t \to t  $ and $ \bar m \to  m $) 
	\begin{equation}
	ds^{2}= -f(r) \, dt^{2}+  k(r, \theta)^{\sigma} \, (\frac{1}{f(r)} \, dr^{2}+ r^{2} d \theta^{2})+ r^{2} \sin^2\theta \, d\phi^{2},\label{eq:bar}
	\end{equation}
	where,
	\begin{align}\label{eq:kbar}
		f(r)&= 1- \frac{2m}{r}+ \frac{ q_{e}^{2}+q_{m}^{2}}{r^{2}},\nonumber \\  k (r,\theta) &= \frac{ {(m^{2}- q_{e}^{2}-q_{m}^{2})  \sin^2\theta}}{{\Delta+(m^{2}- q_{e}^{2}-q_{m}^{2})\sin^2\theta } }, 
	\end{align}
	and 
	\begin{align}
		\Delta=r^{2} -2 \, m \, r + q_{e}^{2}+q_{m}^{2}.
	\end{align}
	Also, the vector potential $A'(r, \theta) $ in   Eq. \eqref{eq:Aprimf} takes the following form
	\begin{equation}
	A'(r, \theta) = A_{0} (r) \, dt + q_{m} \cos \theta \,  d \phi,
	\end{equation}
	where,
	\begin{equation}
	A_{0} (r)= \frac{q_{e}}{r}-\frac{q_{e}}{m(1+p) }, \label{eq:A00}
	\end{equation}
	where   the new parameter $ p $ is defined as
	\begin{equation}
	p=\sqrt{1-\frac{q_{e}^{2}+q_{m}^{2}}{ m }}. 
	\end{equation}
	Also, it is possible to remove the constant term  in Eq.  \eqref{eq:A00}  via a gauge transformation. 
	Using  $ \square \psi (\theta)=0 $ and  metric in Eq. \eqref{eq:bar} the massless scalar field can be found as
	\begin{equation}
	\psi(\theta)=\sqrt{2 \, \sigma } \ln \left(\tan \frac{\theta }{2}\right).	\label{eq:scalar}
	\end{equation}
	
	\section{ A charged Taub–NUT metric  in the presence of a $ \theta  $-dependent  scalar field}\label{4}
	In this section, we aim to use Ehlers transformations and add NUT charge to the   metric in Eq. \eqref{eq:bar}. The electric ansatz of the metric in Eq. \eqref{eq:bar}  is  as follows   ($ q_{m}=0 $)
	\begin{align}
		ds^{2}=- f( dt- \omega \, d\phi)^{2}+  f^{-1} (e^{2 \,  \eta}(d  \rho^{2}+ dz^{2})+ \rho^{2} d\phi^{2}),	\label{eq:LWPcn}
	\end{align}
	where, 
	\begin{align}
		\rho = \sqrt{r^2 -2 \, m \, r+q_{e}^2} \,\, \sin \theta ,\,\,\,\,\,\,\,\,   z=(r-m) \, \cos  \theta, \label{eq:rhobar}
	\end{align}  
	Here, using the  coordinate transformations  in Eq. \eqref{eq:rhobar} in the  metric in
	Eq. \eqref{eq:LWPcn}  and consider the   charged form of the  metric which is given  in Eq. \eqref{eq:bar},  we arrive at
	\begin{align}
		f = 1-\frac{2 \, m}{r}+ \frac{q_{e}^{2}}{r^{2}}, \,\,\,\,\,\,\,\,
		\omega=0,
	\end{align}
	and
	\begin{align}
		e ^{2 \,  \eta}&=\frac{ \Delta \,  (m^{2}-q_{e}^{2})^{\sigma} \sin ^{2\, \sigma} \theta }{(\Delta+ (m^{2}-q_{e}^{2}) \, \sin ^2\theta)^ {\sigma+1}}. \nonumber \\ \Delta&= r^2-2 \, m \, r +q_{e}^{2}.\label{eq:LWPcnut}
	\end{align}
	The coresponding vector potential is defined by  $ A= \frac{q_{e}}{r} dt $. On the other hand, the Ehlers transformations are given by
	\begin{align}
		\mathcal{E}   '=\frac{\mathcal{E} }{1+ i \, c \, \mathcal{E} },\,\,\,\,\,\,\,
		\Phi   '=\frac{\Phi }{1+ i \, c \, \mathcal{E}}, \label{eq:transEhler}
	\end{align}
	where,  $ c $ is a real  parameter. According to    metric in Eq. \eqref{eq:bar}, we define  the gravitational  potential $ \mathcal{E} $, the electro-magnetic potential $\Phi  $ and  $ \chi $  in the following forms ($ q_{m}=0  $)
	\begin{align}
		\mathcal{E}   =1-\frac{2\, m}{r},\,\,\,\,\,\,\,
		\Phi   =\frac{q_{e}}{r}, \, \, \, \, \,  \chi=0.
	\end{align}
	Now, we use the Ehlers transformations in Eq. \eqref{eq:transEhler} and  arrive at the following   transformed Ernst potentials 
	\begin{align}
		\mathcal{E}  '&=\frac{r^2-2 \, m \, r}{r^2+c^2 \, (r-2 \, m)^2}-i \frac{c \, (r-2 \, m)^2}{r^2+c^2 (r-2 \, m)^2},\,\,\,\,\,\,\, \nonumber\\
		\Phi   '&= \frac{q _{e}\, r}{ r^{2}+ c^2 \, (r-2 \, m)^2}- i \frac{ \, c \, q_{e} \, (r-2 \, m)}{r^2+c^2 (r-2 \, m)^2}. \label{eq:Ernst2}
	\end{align}
	Therefore using Eq. \eqref{eq:varepsilon} and  Eq. \eqref{eq:Ernst2} we have
	\begin{align}
		f'&=	\frac{r^2-2 \, m \, r +{q_{e}}^{2}}{r^2+c^2 \, (r-2 \, m)^2},\label{eq:f2}\\
		\chi'&=-\frac{c \, (r-2 \, m)^2}{r^2+c^2 \, (r-2 \, m)^2}. \label{eq:chi2}
	\end{align}
	Now, we consider the following relation and  obtain $ {\omega'} $    
	\begin{align}
		\hat{\phi}\times\nabla\,\chi'=-\rho^{-1}\,f'^2\,\nabla\,\omega'-2\,\hat{\phi}\times Im(\Phi'^\star\,\nabla\,\Phi'),\label{eq:22w}
	\end{align}
	First, using the transformed potential    in Eq. \eqref{eq:Ernst2} we find the imaginary part of  $\Phi^{\star'}\,\nabla\,\Phi ' $  as follows
	\begin{align}
		Im (\Phi^{\star '}\,\nabla\,\Phi ')=	-\frac{2 \, c \, m \, q_{e}^2 \sqrt{\Delta}}{\left(r^2+c^2 \, (r-2 \, m)^2\right)^2 \sqrt{\Delta + (m^{2}-q_{e}^{2})\, \sin^2\theta}},\label{eq:imphi}
	\end{align}
	where, $\Delta $ is defined in Eq. \eqref{eq:LWPcnut} and we have used the  following  transformation  relation for gradient
	\begin{equation}
	\nabla {B}  (r,\theta)= \frac{1}{\sqrt{\Delta + (m^{2}-q_{e}^{2})\, \sin^2\theta }}(\frac{{\partial \, {B} (r,\theta)}}{{\partial \,r}} \sqrt{\Delta} \,\hat r + \frac{{\partial \, {B} (r,\theta)}}{{\partial \,\theta }}\,\hat \theta ),\label{eq:grad2}
	\end{equation}
	then  substituting Eqs. \eqref{eq:rhobar},  \eqref{eq:f2}   \eqref{eq:chi2}, \eqref{eq:imphi} and using the gradient of Eq. \eqref{eq:grad2} in  Eq. \eqref{eq:22w} we obtain 
	\begin{align}
		\omega'(\theta)= 4 \, c \, m \cos\theta. \label{eq:omega}
	\end{align}
	Now, using Eq. \eqref{eq:Ernst2} we have
	\begin{align}
		{A'_t}&= Re(		\Phi   ')= \frac{q _{e}\, r}{ r^{2}+ c^2 \, (r-2 \, m)^2},\,\,\,\,\,\,\nonumber \\ {\tilde A'_\phi}&= Im(	 	\Phi   ')= -\frac{ \, c \, q_{e} \, (r-2 \, m)}{r^2+c^2 (r-2 \, m)^2}.\label{eq:A2}
	\end{align}
	It is known that only using  the Ehlers coordinate  transformations is not enough to find the vector potential. Since   Ehlers transformations rotates the 
	electromagnetic field, we consider  a duality rotation for Ernst potential as follows
	\begin{equation}
	\Phi \to \bar {	\Phi }=  \Phi  \exp {(i \beta)},
	\end{equation}
	where, $ \beta $ is a rotation parameter. Here, we  rotate the components in Eq. \eqref{eq:A2}  as follows
	\begin{equation}
	\left(\begin{matrix} \bar {A_{t}'}\\ \bar {\tilde {A'_{\phi}}}& \end{matrix}\right)=\left(\begin{matrix} \ \cos \beta& -\sin \beta\\ \sin \beta& \cos \beta\end{matrix}\right) \left(\begin{matrix} {A'_t}\\{\tilde A'_\phi}& \end{matrix}\right),\label{eq:Amat}
	\end{equation}
	where ${\tilde {A'_{\phi}}}$  is a twisted potential. Then, using the rotated components  in Eq. \eqref{eq:Amat}   we can find  $ \bar {A_{\phi}'} $ from Eq. \eqref{eq:AtAphi}. Therefore,  we  arrive at the components of the vector potential as  follows
	\begin{align}
		\bar {A_{t}'} (r)&=\frac{q_{e} \,  \,\left( r \, \cos \beta + c \, (r-2\, m) \,\sin \beta \right) }{r^{2}+ c^{2} (r-2\,m)^{2}},\nonumber \\ 
		\bar {A_{\phi}'}(r, \theta)&=  \, q_{e} \cos \theta \, (c  \, \cos \beta -\sin \beta) - \omega' \, \bar A_{t}',\label{eq:A3}
	\end{align} 
	It should be noted that the Ehlers transformation does not change the function $ \eta $ therefore $ \tilde \eta=\tilde\eta'$.
	Then by using Eqs. \eqref{eq:LWPcn}, \eqref{eq:rhobar}, \eqref{eq:LWPcnut}, \eqref{eq:f2} and  Eq. \eqref{eq:omega} the metric is rewritten as 
	\begin{align}
		ds'^{2} &= -\frac{\Delta}{r^2+c^2 (r-2 \, m)^2}(dt- 4 \, c \, m \cos\theta \, d\phi )^{2}+ \frac{r^2+c^2 (r-2 \, m)^2}{\Delta}\left(\frac{(m^2-{q_{e}}^{2}) \sin ^2\theta }{  \Delta+(m^2-{q_{e}}^{2})\sin ^2\theta  }\right)^{\sigma } \nonumber\\ &\times (dr^{2} +\Delta \, d \theta^{2})+
		({r^2+c^2 (r-2 \, m)^2} )\sin ^2\theta \, d \phi^{2},\label{eq:nutm}
	\end{align}
	where  $\Delta $ is defined in Eq. \eqref{eq:LWPcnut}. Now, we use the following coordinate transformations and redefine the parameters in Eq. \eqref{eq:A3}  and Eq. \eqref{eq:nutm}      in the following way
	\begin{align}
		\tilde r &= r \sqrt{1+c^{2}}-\frac{2 \, m \, c^{2}}{\sqrt{1+c^{2}}},\,\,\,\,\,\,\,\,\,\,\, \tilde t=\frac{t}{ \sqrt{1+c^{2}}},\,\,\,\,\,\,\,\,\tilde{q}=q_{e}{\sqrt{1+c^2}}, \nonumber \\\nonumber \\
		m &=-\frac{n }{2 \, c}\sqrt{1+c^{2}},\,\,\,\,\,\,\,\,\,\,\,\,\,\,\,\,\,\,\,\,\,\,\,\,\,\, c =\frac{\tilde m-\sqrt{\tilde m^{2}+n^{2}}}{n},  \,\,\,\,\,\,\,\, \cos \beta = \frac{1}{\sqrt{1+c^{2}}}. \label{eq:Ehlers}
	\end{align}
	where, $ n $ is the  NUT charge.    Now, we remove   tilde from parameters   ($ \tilde r \to r, \,\, \tilde t \to t  $, $ \tilde{q} \to  q,  \tilde m  \to m $) and derive the final  form of the class of charged Taub-NUT metrics in the presence of a $ \theta $-dependent scalar field as follows 
	\begin{align}
		ds'^{2} &= -f(r) \, (dt+ 2 \, n \cos \theta \, d \phi)^{2}+ { k (r, \theta)}^{\sigma} \, ( \frac{1} {f(r)} dr^{2}+ (r^2 +n ^2) \, d\theta^{2}) \nonumber\\ &+ (r^2+n^{2}) \, \sin ^2\theta \, d \phi^{2},\label{eq:qnut}
	\end{align}
	where,
	\begin{align}\label{eq:kTaub}
		f(r)&=\frac{\Delta_{n}}{r^2+n ^2}, \\\nonumber
		k (r,\theta)&= \frac{  \left(m^2+n^2-q^{2}\right) \sin ^2\theta }{  \Delta_{n}+ \left(m^2+n^2-q^{2}\right)\sin ^2\theta },
	\end{align}
	and
	\begin{align}
		\Delta _{n}=r^2-2 \, m \, r +q^{2}-n^2.
	\end{align}
	The  specific value for $ \beta $ in Eq. \eqref{eq:Ehlers} is chosen in a way that we arrive at  $ 	\bar {A_{\phi}'}=- \omega' 	\bar {A_{t}'}  $ in Eq. \eqref{eq:A3}. Therefore  the new vector potential is  as below
	\begin{equation}
	A' (r, \theta) =  {A_{t}'} \, dt+  {A_{\phi}'} \, d\phi= \frac{q \, r}{r^{2}+ n^{2}} (dt + 2 \, n  \cos \theta \, d\phi),
	\end{equation}
	where, we  have removed bars ($ \bar {A_{t}'} \to  {A_{t}'} $, $ \bar {A_{\Phi}'} \to  {A_{\Phi}'} $).
	Also, using  $ \square \psi (\theta)=0 $ and the charged  Taub-NUT metric in Eq. \eqref{eq:qnut}, the massless scalar field  can be found which is the same as	 Eq. \eqref{eq:scalar}. It should be noted that 
	it is also possible to add  NUT parameter   to the seed metrics  in Eq. \eqref{eq:metric} via Ehlers transformations and then obtain charged Taub-NUT metrics   using Harrison's transformations.   It should be noted that the commutativity of the Ehlers and Harison transformation is established  when  those transformations applied to the same LWP metric. This subtle point was extensively discussed in \cite{barrientos2024mixing}.  
	
	The  Taub-NUT metric  which is given by Eq. \eqref{eq:qnut}   $ (q=0) $ can be  transformed to the following form of  the  LWP metric
	\begin{align}
		ds^{2}=- f( dt- \omega \, d\phi)^{2}+  f^{-1} (e^{2 \,  \eta}(d  \rho^{2}+ dz^{2}) + \rho^{2} d\phi^{2}),	\label{eq:LW2}
	\end{align}  
	by using   Weyl's coordinates as follows
	\begin{align}
		f&=\frac{({R_{+}}+{R_{-}})^2-4 \left(m^2+n^2\right)}{({R_{+}}+{R_{-}}+2 \, m)^2+4 \, n^2}, \\
		e^{2 \, \eta}&=\frac{({R_{+}}+{R_{-}})^2-4 \left(m^2+n^2\right)}{4 \, {R_{+}}{R_{-}}} \left(\frac{4 \, (m^2+n^2)-({R_{+}}-{R_{-}})^2}{4 \, {R_{+}}{R_{-}}}\right)^{\sigma}, \\
		\omega&=-\frac{n}{\sqrt{m^{2}+n^{2}}}(R_{+}-R_{-}),\\
		R_{\pm}^{2}&=\rho^{2} +(z \pm \sqrt{m^{2}+n^{2}})^{2},
	\end{align}
	and also  the following relations are useful 
	\begin{align}
		\rho&=\sqrt{r^{2}-2 \, m\, r-n^{2}} \sin \theta, \, \, \, \, \, \, \, \,\, \, \, \,\, \, \,
		z= (r-m)\, \cos \theta,  \nonumber\\
		R_{\pm}&=r-m \pm \sqrt{m^2+n^2} \, \cos \theta .\label{eq:rowe}
	\end{align} 
	\subsection{Curvature singularities}
	Here, we derive the singularties of the metric  via the Ricci scalar. The Ricci scalar of the metric in Eq. \eqref{eq:qnut} is given by
	\begin{equation}
	R=\frac{2 \, \sigma }{(r^2+n^{2})  \sin ^2 \theta  } \left(\frac{ {\Delta_{n}+(m^{2}+n ^{2}- q^{2})\sin^2\theta}}{{ (m^{2}+n^{2}- q^{2})  \sin^2\theta} } \right) ^{\sigma},	\label{eq:Ricci}
	\end{equation}
	At singular points   the Ricci scalar becomes infinite. According to Eq. \eqref{eq:Ricci}  in order to avoid singular points at $ r \to \infty $, the parameter $ \sigma  $ should be  $ \sigma \le1 $. For $ \sigma < 0 $, the singular points are as follows
	\begin{equation}
	r= m \pm  \cos \theta\sqrt{ m^{2} + n^{2}-q^{2}},
	\end{equation}
	Therefore, we choose    $ 0 \le \sigma  \le 1  $ and  
	singularities are located only at  $ \theta=0 , \, \pi $. The Taub-NUT metrics also have  singularities at the points $ \theta =0 $ and $ \theta =\pi $ due to the change of  metric signature at these points  \cite{bonnor1969new}. Therefore, the singularities of the generated charged-Taub-NUT metrics incorporating a $\theta$-dependent scalar field are exactly coincided  with the Taub-NUT black holes. The generated metric in Eq. \eqref{eq:qnut}   can be considered  as  a black hole with two axial singularities.    It should be noted that the same singularities appears in pure Taub-NUT black holes.   
	
	\section{Gravitational lensing}\label{5}
	In this section, we will investigate the gravitational lensing for the class of  metrics in Eq. \eqref{eq:qnut}. We calculate the deflection angle of light by  using the approach of    Gibbons and Werner which is based on the Gauss-Bonnet theorem \cite{gibbons2008applications}. The optical metric in the equatorial plane of space-time is given by
	\begin{equation}\label{eq:49}
	d t^2 = \tilde{g}_{r r} \, d r^2 + \tilde{g}_{\phi \phi} \, d\phi^2,
	\end{equation}
	where,
	\begin{equation}\label{eq:50}
	\begin{aligned}
	& \tilde{g}_{r r} ={{ {k(r,\theta)}}^{\sigma }}{f(r) ^{-2}} ,\\
	& \tilde{g}_{\phi \phi} = (r^{2}+n^{2}) {f(r)^{-1}},
	\end{aligned}
	\end{equation}
	and the  definitions of $ f(r) $ and $ k(r,\theta) $  can be found in Eq. \eqref{eq:kTaub}. It is known that the equation of Gaussian curvature is in the following form
	\begin{equation}\label{eq:51}
	\kappa = - \, \frac{1}{\sqrt{\tilde{g}}} \,\left( \partial_r \, \big( \frac{1}{\sqrt{\tilde{g}_{r r}}} \, \partial_r \, \sqrt{\tilde{g}_{\phi \phi}} \big) + \partial_\phi \, \big( \frac{1}{\sqrt{\tilde{g}_{\phi \phi}}} \, \partial_r \, \sqrt{\tilde{g}_{r r}} \big)\right ),
	\end{equation}
	where  $ \tilde{g} $ is the determinant of the metric. The deflection angle of gravitational lensing can be obtained by the following relation
	\begin{equation}\label{eq:53}
	\alpha = - \, \int_{0}^{\pi} \, \int_{r_{0}}^{\infty} \, \kappa \, \sqrt{\tilde{g}} \, d r \, d \phi,
	\end{equation}
	where, $ r_{0} $ denotes  the minimum distance from the source and can be determined via  the null geodesics equation.  Now, using  Eqs. 
	\eqref{eq:49}, \eqref{eq:50}, and   Eq. \eqref{eq:51}   in Eq. \eqref{eq:53}
	we arrive at
	\begin{equation}
	\alpha = - \, \int_{0}^{\pi} \, \int_{r_0}^{\infty} \, \partial_r \,\left(  {f(r)^{\frac{1}{2}}} k(r,\theta)^{-\frac{\sigma }{2}} (r^{2}+n^{2})^{\frac{-1}{2}} (r- \frac{(r^{2}+n^{2})}{2} \, \frac{ f'(r)}{ f(r)  })  \right) \, d r \, d \phi.\label{eq:del}
	\end{equation}
		\begin{figure}[t]
		\centering
		\subfloat[\label{subfig:m}]	{\includegraphics[width=0.45\textwidth]{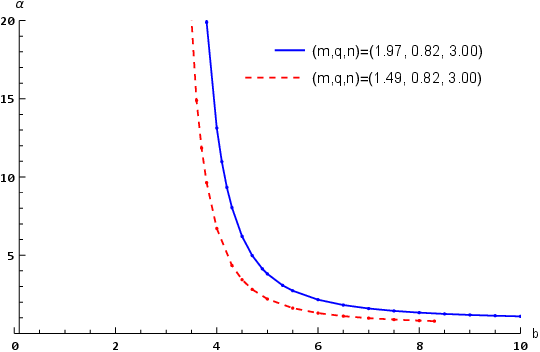}}
		\qquad
		\subfloat[\label{subfig:q}]{\includegraphics[width=0.45\textwidth]{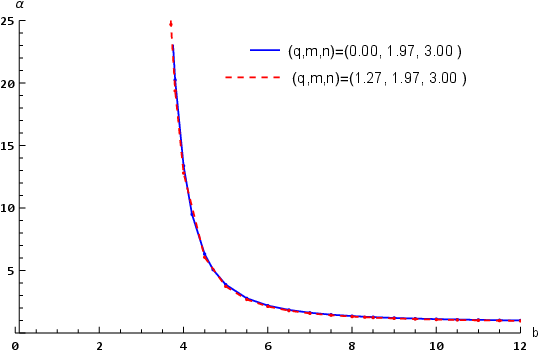}}
		\qquad
		\subfloat[a][\label{subfig:nut}]{\includegraphics[width=0.45\textwidth]{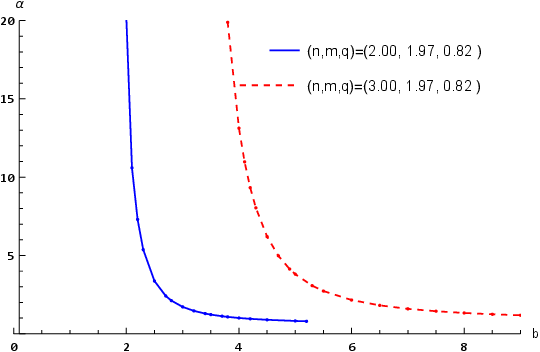}}
		\qquad
		\subfloat[\label{subfig:sigma}]{\includegraphics[width=0.45\textwidth]{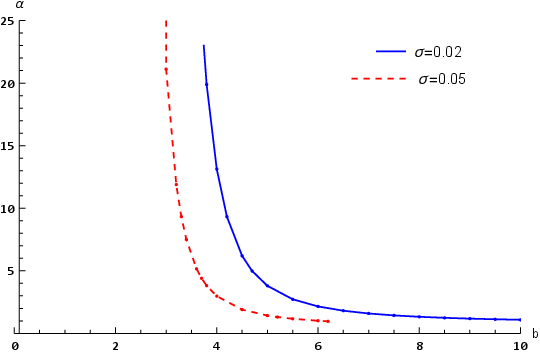}}		
		\caption{ \normalfont The   deflection angle $ \alpha $ as a function of impact parameter $ b $. (a): We have considered $ \sigma=0.02 $, $ n=3.00 $ and $ q=0.82 $ with two different values of mass i.e.  $ m= 1.97 $ \, (solid blue curve) and  $1.49 \, $  (dashed red curve).   (b):   $ \sigma=0.02 $, $ m=1.97 $  and $ n=3.00 $ for $ q=0.00 $ ( solid blue curve) and  $ q=1.27 $ (dashed red curve). (c): $ \sigma=0.02 $, $ m=1.97 $ and  $ q=0.82 $ for $ n=2.00 $ (solid blue curve) and $ n=3.00 $ (dashed red curve). (d):  $ m=1.97 $, $ n=3.00 $ and  $ q=0.82 $ for $ \sigma=0.02 $  (solid blue curve)  and $ \sigma=0.05 $ (dashed red curve). }
		\label{fig:lensing}	
	\end{figure}
	To derive the deflection angle $ \alpha $,
	expanding  the whole  integrand  up to  the order of $  r^{-2} $, ($0 <  \sigma   < 1$ ) we have 
	\begin{equation}
	\alpha = - \, \int_{0}^{\pi} \, 	\left(\frac{m^2+n ^2-q_{}^2}{r_{0}^2}\right)^{-\frac{\sigma }{2}} \left(1-\frac{m \, (\sigma +2)}{r_{0}}+\frac{m^2 (\sigma \, (\sigma +3)-3)+3 \, {q_{}}^2-7 \,n ^2}{2 \, r_{0}^2}\right) \, d \phi. \label{eq:alpha}
	\end{equation}
	In order to obtain $ r_{0} $  we  consider the charged Taub-NUT metrics in Eq. \eqref{eq:qnut}    at $ \theta=\frac{\pi}{2} $ as follows      
	\begin{equation}
	ds^{2}= - f(r) \, dt^{2}+\frac{{k(r,\theta)}^\sigma}{f(r)} \, dr^{2}+{(r^{2}+n^{2})} \, d\phi^{2}. \label{eq:qnuttheta}
	\end{equation}
	Now assuming the three dimensional spacetime in Eq. \eqref{eq:qnuttheta}, the geodesic Lagrangian is given by
	\begin{equation}
	\mathcal{L}= -\frac{1}{2} \, f(r) \, \dot{t}^{2}+ \frac{1}{2}\frac{{k(r,\theta)}^\sigma}{f(r)} \, \dot{r}^{2}+ \frac{1}{2} {(r^{2}+n^{2})} \, \dot{\phi}^{2},
	\end{equation}
	where, dot denotes the $ \frac{d}{d\lambda} $ and $ \lambda  $ is the affine parameter which leads to the following equations of motion
	\begin{align}
		f(r) \, \frac{d t}{d \lambda} &= E = const,\label{eq:Ef}\\
		{(r^{2}+n^{2})} \, \frac{d \phi}{d \lambda} & = L = const.\label{eq:Lf}
	\end{align}
	Therefore by using both  Eq. \eqref{eq:Ef} and Eq. \eqref{eq:Lf} we arrive at
	\begin{equation}
	\frac{f(r)}{r^{2}+n^{2}} \, \frac{d t}{d \phi} = \frac{E}{L} = \frac{1}{b}, \label{eq:b}
	\end{equation}
	where, $ b $ is a constant. Then, using  Eq. \eqref{eq:qnuttheta}, $(\frac{dr}{ d \lambda})^2 $ takes the following form
	\begin{equation}
	\frac{{k(r,\theta)}^\sigma}{f(r)} \big( \frac{d r}{d \lambda} \big)^2= f(r)\big( \frac{d t}{d \lambda} \big)^2-{(r^{2}+n^{2})} \big( \frac{d \phi}{d \lambda} \big)^2.\label{eq:drd2}
	\end{equation}
	Now, using $ d \lambda \to d\phi$  in Eq. \eqref{eq:drd2}   we have
	\begin{equation}
	\frac{{k(r,\theta)}^\sigma}{f(r)} \big( \frac{d r}{d \phi} \big)^2=  \frac{(r^{2}+ n^2)^2}{   b^{2} \, f(r)} -  ({r^{2}+n^{2}}),\label{eq:drdl}
	\end{equation}
	where, we  used Eq. \eqref{eq:b}. Then, we use  $ 	r= \frac{1}{u} $, therefore,   Eq. \eqref{eq:drdl} can be written as
	\begin{equation}
	\big( \frac{d u}{d \phi} \big)^2 = \frac{f(r) \, (1+n^{2} u^{2})}{ \, {k(r,\theta)}^{\sigma}}\left(\frac{1+n^{2} u^{2}}{ b^{2} \, f(r)}- u^{2}\right). \label{eq:du}
	\end{equation}
	Then,      expanding  Eq. \eqref{eq:du}  up to  $ u^{2} $ and the first order of $ \sigma $  we  have
	\begin{align}\label{eq:nu}
		\frac{d^2 u}{d \phi^2} + u &=3 \, m \, u^2+\frac{2  \, n ^2 \, u}{b^2}+\sigma  [-\frac{2 \, n ^2 u \, \ln \left(u^2 \left(m^2+n ^2-{q}^2\right)\right)+4 \, n ^2 m \, u^2 + m  }{b^2}-\frac{1}{b^2 u}\\ \nonumber  &-3 \, m \, u^2 \, \ln \left(u^2 \left(m^2+n ^2-{q}^2\right)\right)+u \, \ln \left(u^2 \left(m^2+n ^2-{q}^2\right)\right)+2 \, m \, u^2].
	\end{align}
Then, we  solve Eq. \eqref{eq:nu} numerically, where its solutions determines $r_{0}$ ($u=\frac{1}{r_{0}}$).  Now, we insert obtained  values for  $ r_{0}   $     in  Eq. \eqref{eq:alpha} and calculate the deflection angle $ \alpha $ of light.
	The   deflection angle $ \alpha $ of light   as a function of $ b $ is depicted in  Fig. \ref{fig:lensing}. Fig. \ref{fig:lensing} contains four graphs that show that as  the impact parameter $ b $ increases the deflection angle  $ \alpha $ of light decreases. Also,  Fig. \subref* {subfig:q} shows  that variation of the electric charge $ q $ has little effect on behavior of deflection angle $ \alpha $. 
	\section{Quasi normal modes}\label{6}
	In the following, we are going to calculate the QNMs of  metric in Eq. \eqref{eq:qnut}. For  this purpose, we use the light ring method in the eikonal limit 
	\cite{ferrari1984oscillations, ferrari1984new, mashhoon1985stability}. There is no need to separate wave equations in the light ring method. Particular QNMs can be  obtained   by using  this method.   We investigate the characteristics of unstable circular photon   orbits in the equatorial plane.
	Therefore, according to light ring method,  one must  choose $ \theta=\frac{\pi}{2} $  in  the metric in Eq. \eqref{eq:qnut}.  Also, we  expand the metric components up to the first order of $\sigma$, and the second orders of $n$ and $q$, and omit the terms $\sigma\, n $, $\sigma\, q$, $n \, q^2$ and $n \, q$. We apply these approximations in  the whole of the calculations. Therefore, after expanding components of metric in Eq. \eqref{eq:qnut}, we have
	\begin{equation}
	\begin{aligned}
	d s^2 & = -\left(1-\frac{2\,m}{r}+\frac{q^2}{r^2}-\frac{2\,n^2}{r^2}\,(1-\frac{m}{r})\right)\,dt^2+\frac{1}{r\,(1-2\,m/r)}\, \\ 	&\times \left(r-\frac{q^2}{r\,(1-2\,m/r)}
	+\frac{2\,n^2}{r\,(1-2\,m/r)}\,(1-\frac{m}{r})+r\,\sigma\,\ln\big(\frac{m^2}{r^2-2\,m\,r+m^2}\big)\right) \,dr^2  \\
	&+\left(r^2+n^2+r^2\,\sigma\,\ln\big(\frac{m^2}{r^2-2\,m\,r+m^2}\big)\right)\,
	d\theta^2+ \left(r^2+n^2\right)\,d\phi^2.
	\end{aligned}\label{eq:metricff}
	\end{equation}
	We consider the QNMs as $ Q = \Omega + i \, \Gamma $. In the light ring method, $ \Omega $ can be written as follows
	\begin{equation}\label{q2_label}
	\Omega=\pm\,j\,\Omega_{\pm}.	
	\end{equation}
	Also, we can represent massless wave perturbations as superpositions of the following eigenmodes  
	\begin{equation}\label{51_label}
	e^{i \, (\Omega \, t - \iota \, \phi)} \, S_{\Omega \, j \, \iota \, s} (r, \theta),
	\end{equation}
	where, $ \Omega $ is wave's frequency and $ S $ is the wave's spin. $ \iota $ and $ j $ are angular momentum and the following condition exists between them
	\begin{equation}\label{qnms2_label}
	\lvert \iota \rvert \le j.
	\end{equation}
	Also, the following two conditions exist in the eikonal limit
	\begin{equation}\label{qnms3_label}
	\Omega \gg 1/M,\qquad\lvert \iota \rvert = j \gg 1.
	\end{equation}
	In the eikonal limit, the frequency of perturbations can be represented by the following equation
	\begin{equation}\label{qnms4_label}
	\Omega = \iota \, \frac{d \phi}{d t} = \pm \, j \, \Omega_{\pm}.
	\end{equation} 
	To calculate QNMs, we must first calculate the radius of unstable null circular orbits, for this we use the following two equations
	\begin{equation}\label{54_label}
	g_{t t}+ g_{\phi \phi} \, \big( \frac{d \phi}{d t} \big)^2 = 0,
	\end{equation}
	and
	\begin{equation}\label{55_label}
	\Gamma_{t t}^r + \Gamma_{\phi \phi}^r  \, \big( \frac{d \phi}{d t} \big)^2 = 0.
	\end{equation}
	Eq. \eqref{54_label} is written for each null path and Eq. \eqref{55_label} shows the radial component of the geodesic equation. By using  Eq. \eqref{eq:metricff}, we can write Eq. \eqref{54_label} and Eq. \eqref{55_label} as below
	\begin{equation}\label{56_label}
	\begin{aligned}
	\frac{d\phi}{dt}=\frac{r\,(2\,r^2+q^2-3\, n^2)+4\,m\,(n^2-r^2)}{2\,r^4\,\sqrt{(1-2\,m/r)}},
	\end{aligned}
	\end{equation}
	\begin{equation}\label{57_label}
	\begin{aligned}
	\frac{d\phi}{dt}=\frac{r\,(2\,m\,r-q^2)+n^2\,(2\,r-3\,m)}{2\,r^4\,\sqrt{m/r}}.
	\end{aligned}
	\end{equation}
	We put $ r $ as $ 3 \, m + \bar{\epsilon} $ in Eq. \eqref{56_label} and Eq. \eqref{57_label}, where $3\,m$ represents the radius of the light ring of the Schwarzschild black hole and $ \bar{\epsilon} $ is the perturbation parameter. Then, as the right side of Eq. \eqref{56_label} and Eq. \eqref{57_label} are equal,  we have
	\begin{equation}\label{58_label}
	\bar{\epsilon}=-\frac{12\,m\,(3\,q^2-4\,n^2)}{54\,m^2-39\,q^2+41\,n^2},\quad\Rightarrow\quad r_0=3\,m-\frac{12\,m\,(3\,q^2-4\, n^2)}{54\,m^2-39\,q^2+41\, n^2}.
	\end{equation}
	By replacing $ r_0 $ in Eq. \eqref{56_label} or Eq. \eqref{57_label}, we arrive at
	\begin{equation}\label{60_label}
	\Omega_\pm = \frac{d \phi}{d t} = \frac{1}{3\,\sqrt{3}\,m}\pm\frac{1}{18\,\sqrt{3}\,m^3}\,\big(q^2-\frac{5\, n^2}{3}\big).
	\end{equation} 
	To calculate the term $ \Gamma $ in QNMs, we perturb the coordinates of the circular orbit \\ $ \bar{x}^\mu = (t, r, \theta, \phi) = (t, r_0, \frac{\pi}{2}, \Omega^\prime t) $ as follows
	\begin{equation}\label{61_label}
	r = r_0 \, [1 + \epsilon_0 \, \tau_r(t)], \quad \phi = \Omega_\pm \, [t + \epsilon_0 \, \tau_\phi (t)], \quad \ell = t + \epsilon_0 \, \tau_\ell(t),
	\end{equation}
	where, $ \epsilon_0 $ is a small perturbation and the conditions for $ \tau_r(t) $ and $ \tau_\ell(t) $ are as follows
	\begin{equation}\label{62_label}
	\tau_r(0) = \tau_\ell(0) = 0.
	\end{equation}
	Also, according to Eqs. \eqref{qnms4_label} and Eq. \eqref{61_label}, we have $ \tau_\phi(t) = 0 $. We consider the perturbed propagation vector as follows 
	\begin{equation}\label{q11_label}
	K^\mu = \frac{d x^\mu}{d \ell} = \big(1 - \epsilon_0 \, \frac{\partial\,\tau_\ell}{\partial\,t}, \, \epsilon_0 \, r_0 \, \frac{\partial\,\tau_r}{\partial\,t}, \, 0, \, \Omega_\pm \, (1 - \epsilon_0 \, \frac{\partial\,\tau_\ell}{\partial\,t})\big).
	\end{equation}
	We can write the conservation law for the  congruence of  null rays as below
	\begin{equation}\label{q12_label}
	\nabla_\mu \, (\varrho \, K^\mu) = 0,
	\end{equation}
	where $ \varrho $ is the density of null rays. Also, using Eq.  \eqref{q12_label}, we have
	\begin{equation}\label{q13_label}
	\frac{1}{\varrho} \, \frac{d \varrho}{d \ell} = - \nabla_\mu K^\mu = - \frac{1}{\sqrt{- g}} \, \frac{\partial}{\partial x^\alpha} \, (\sqrt{- g} \, K^\alpha),
	\end{equation}
	Then, by using Eq.  \eqref{q11_label} we can write   Eq.  \eqref{q13_label}   as follows
	\begin{align}
		\frac{1}{\varrho} \, \frac{d \varrho}{d \ell} &=-\frac{1}{\sqrt{-g}} \frac{\partial}{\partial t} (\sqrt{-g} \, (1-\epsilon_0 \, \tau_{\ell}'))-\frac{1}{\sqrt{-g}} \frac{\partial}{\partial r} (\sqrt{-g} \, \epsilon_0 \, r_{0} \, \tau_{r}') \nonumber \\&-  \frac{\Omega_\pm}{\sqrt{-g}} \frac{\partial}{\partial \phi} (\sqrt{-g} (1-\epsilon_0 \, \tau_{\ell}')),\label{q1d_label}
	\end{align} 
	where prime denotes the derivative with respect to the parameter $ t $.  Using the metric in Eq. \eqref{eq:metricff}   we have
	\begin{align}
		\sqrt{-g}= r^{2}+n^{2} +\sigma \,  r^{2} \ln \left(\frac{m^2}{r^2-2 \, m \,  r+m^2}\right).
	\end{align}  
	By using Eq. \eqref{58_label}  and Eq. \eqref{q11_label}  $ \sqrt{-g} $ is rewritten as follows
	\begin{align}
		\sqrt{-g}= \frac{19 \, n ^2}{3}+9 \, m^2 (1+ 2 \,\epsilon_0 \, \tau_r \,  -\sigma  \ln 4)-4 \, q^2 (1+ 2 \,\epsilon_0 \,\tau_r   ) .\label{q1e_label}
	\end{align}
	Therefore, using  Eq. \eqref{q1d_label} and Eq. \eqref{q1e_label} we arrive at
	\begin{equation}
	\frac{1}{\varrho} \, \frac{d \varrho}{d \ell} = - \epsilon_0 \left(\frac{\partial (2\, \tau_r - \tau_{\ell}' )}{\partial t}+ \frac{\partial (r_{0 } \, \tau_r')}{\partial r}+ \Omega_\pm \frac{\partial (2 \, \tau_r - \tau_{\ell}')}{\partial \phi} \right).
	\end{equation} 
	Therefore   by using Eq. \eqref{61_label}  we have  the following relation
	\begin{align}
		\frac{1}{\varrho} \, \frac{d \varrho}{d t} &=\frac{d(\epsilon_0 \, r_{0} \, \tau_{r}')}{dt} \, \frac{dt}{dr}+ O(\epsilon_0) \nonumber \\ &=-\frac{\tau_r^{\prime \prime}(t)}{\tau_r^\prime(t)}+ O(\epsilon_0),\label{q1b_label}
	\end{align} 
	where we used Eq.  \eqref{q11_label} ($ \frac{dr}{dt}= \epsilon_0 \, r_{0} \, \tau_{r}'  $).  Now, we first derive the $\tau_r $. By writing the radial component of the geodesic equation, one has
	\begin{equation}\label{q15_label}
	\frac{d^2 r}{d \ell^2} + \Gamma_{t t}^r  \,\big( \frac{d t}{d \ell} \big)^2 + \Gamma_{\phi \phi}^r \, \big( \frac{d \phi}{d \ell} \big)^2 + \Gamma_{\theta \theta}^r \, \big( \frac{d \theta}{d \ell} \big)^2 = 0.
	\end{equation}
	By using Eq.\eqref{eq:metricff} and expanding Eq. \eqref{q15_label} up to first order of $ \epsilon_0 $, we have
	\begin{equation}\label{q16_label}
	\frac{\epsilon_0}{81\,m^3}\,\left(9\,m^2\,\big(27\,m^2-6\,q^2+8\, n^2\big)\,\tau_r^{\prime \prime}(t)+\big(q^2+n^2-9\,m^2\,(1+\sigma\,\ln4)\big)\,\tau_r(t)\right)=0.
	\end{equation}
	$ \tau_r(t) $ can be written as below
	\begin{equation}\label{q17_label}
	\tau_r(t) = \sinh (\zeta \, t),
	\end{equation}
	where $ \zeta $ is given by
	\begin{equation}\label{q18_label}
	\zeta = \frac{1}{3\,\sqrt{3}\,m}\,\left(1+\frac{q^2}{18\,m^2}-\frac{11\,n^2}{54\,m^2}+\sigma\,\ln2\right).
	\end{equation}
	According to Eqs. \eqref{q1b_label} and Eq. \eqref{q17_label}, $ \varrho(t) $ can be written as follows
	\begin{equation}\label{q19_label}
	\varrho(t) = \varrho(0) \, \frac{1}{\cosh(\zeta)} \simeq 2 \, \varrho(0) \, (e^{- \zeta \, t} - e^{- 3 \, \zeta \, t} + e^{- 5 \, \zeta \, t} - \dots).
	\end{equation}
	Also, in the light ring method, we can represent $ \Gamma $ as below
	\begin{equation}\label{q20_label}
	\Gamma = \big(\mathcal{N} + \frac{1}{2} \big) \, \zeta.
	\end{equation}
	Therefore, QNMs for metric Eq. \eqref{eq:metricff} are equal to
	\begin{equation}\label{q21_label}
	\begin{aligned}
	Q & = \Omega + i \, \Gamma\\
	& = j \, \Big\{\frac{1}{3\,\sqrt{3}\,m}\pm\frac{1}{18\,\sqrt{3}\,m^3}\,\big(q^2-\frac{5\, n^2}{3}\big)\Big\} + i \, \big(\mathcal{N} + \frac{1}{2} \big) \, \Big\{\frac{1}{3\,\sqrt{3}\,m}\,(1+\frac{q^2}{18\,m^2}-\frac{11\, n^2}{54\,m^2}\\
	&+\sigma\,\ln2)\Big\}.
	\end{aligned}
	\end{equation}
	Eq. \eqref{q21_label} for $ q =n=\sigma= 0 $ represents the QNMs of Schwarzschild black holes. The presence of the scalar field adds a positive value  $ 0.70 \, \sigma $ to the imaginary part of  QNMs.  
	\section{Conclusion}\label{7}
	In this paper, we considered a class of axially symmetric metrics coupled with a $\theta$-dependent scalar field  and adding charge and NUT parameter via Harison and Ehlers transformations.  The generated charged-Taub-NUT-scalar metrics  represent exact solutions to Einstein's equations. Since the scalar field equation is decoupled  from the gravitational and the Maxwell field equations, the Ernst scheme is utilized without any modifications.  Then, we obtained the singularities of the curvature for generated charged-Taub-NUT-scalar metrics. The generated metrics   have   axial singularities at the points $\theta=0$ and $\theta=\pi$. Therefore, the singularities of the generated metrics are coincided with the Taub-NUT metrics  that have   singularities at  $\theta=0$ and $\theta=\pi$ due to the  change of   metric signature at these points.   We investigate the gravitational lensing of the class of metrics and obtained the deflection angle of light for the class of  metrics. Finally, we derived QNMs of the metric  via the light-ring method.  We hope to  derive  the rotating form of the charged-Taub-NUT-scalar metrics and  study other  aspects of the class of metrics in the  future.
	
	\section*{Acknowledgements}
	We would like to thank Isfahan University of Technology for financial support that made available to us.
	\bibliographystyle{elsarticle-num}
	\bibliography{abibliography}{}

\begin{thebibliography}{10}
\expandafter\ifx\csname url\endcsname\relax
  \def\url#1{\texttt{#1}}\fi
\expandafter\ifx\csname urlprefix\endcsname\relax\def\urlprefix{URL }\fi
\expandafter\ifx\csname href\endcsname\relax
  \def\href#1#2{#2} \def\path#1{#1}\fi

\bibitem{hawking1972black}
S.~W. Hawking,
  \href{https://link.springer.com/article/10.1007/bf01877517}{Black holes in
  general relativity}, Communications in Mathematical Physics 25 (1972)
  152--166.
\newline\urlprefix\url{https://link.springer.com/article/10.1007/bf01877517}

\bibitem{novikov1973astrophysics}
I.~D. Novikov, K.~S. Thorne,
  \href{https://books.google.com/books?hl=en&lr=&id=sUr-EVqZLckC&oi=fnd&pg=PA343&dq=black+holes&ots=aJVNnmt2pf&sig=L1JUN5tXDpFyPicQudOrB5CoKD8#v=onepage&q=black%20holes&f=false}{Astrophysics
  of black holes}, Black holes (Les astres occlus) 1 (1973) 343--450.
\newline\urlprefix\url{https://books.google.com/books?hl=en&lr=&id=sUr-EVqZLckC&oi=fnd&pg=PA343&dq=black+holes&ots=aJVNnmt2pf&sig=L1JUN5tXDpFyPicQudOrB5CoKD8#v=onepage&q=black%20holes&f=false}

\bibitem{carr1974black}
B.~J. Carr, S.~W. Hawking,
  \href{https://academic.oup.com/mnras/article-abstract/168/2/399/2604878}{Black
  holes in the early universe}, Monthly Notices of the Royal Astronomical
  Society 168~(2) (1974) 399--415.
\newline\urlprefix\url{https://academic.oup.com/mnras/article-abstract/168/2/399/2604878}

\bibitem{preskill1992black}
J.~Preskill,
  \href{https://www.worldscientific.com/doi/pdf/10.1142/9789814536752#page=37}{Do
  black holes destroy information}, in: Proceedings of the International
  Symposium on Black Holes, Membranes, Wormholes and Superstrings, S. Kalara
  and DV Nanopoulos, eds.(World Scientific, Singapore, 1993) pp, World
  Scientific, 1992, pp. 22--39.
\newline\urlprefix\url{https://www.worldscientific.com/doi/pdf/10.1142/9789814536752#page=37}

\bibitem{taylor2000exploring}
E.~F. Taylor, J.~A. Wheeler,
  \href{https://www.eftaylor.com/exploringblackholes/AAAAREADME180423v1.pdf}{Exploring
  black holes}, Vol.~98, Addison Wesley Longman San Francisco, 2000.
\newline\urlprefix\url{https://www.eftaylor.com/exploringblackholes/AAAAREADME180423v1.pdf}

\bibitem{alexander2012drives}
D.~M. Alexander, R.~C. Hickox,
  \href{https://www.sciencedirect.com/science/article/abs/pii/S1387647311000583}{What
  drives the growth of black holes?}, New Astronomy Reviews 56~(4) (2012)
  93--121.
\newblock \href {http://arxiv.org/abs/1112.1949} {\path{arXiv:1112.1949}}.
\newline\urlprefix\url{https://www.sciencedirect.com/science/article/abs/pii/S1387647311000583}

\bibitem{novikov2013physics}
I.~Novikov, V.~Frolov,
  \href{https://books.google.com/books?hl=en&lr=&id=Q4EiCQAAQBAJ&oi=fnd&pg=PR8&dq=black+holes&ots=c13dqLzRRY&sig=Fw1Yj-6DYEIMrvkNX0Ai_sJBK00}{Physics
  of black holes}, Vol.~27, Springer Science \& Business Media, 2013.
\newline\urlprefix\url{https://books.google.com/books?hl=en&lr=&id=Q4EiCQAAQBAJ&oi=fnd&pg=PR8&dq=black+holes&ots=c13dqLzRRY&sig=Fw1Yj-6DYEIMrvkNX0Ai_sJBK00}

\bibitem{penrose1973naked}
R.~Penrose,
  \href{https://nyaspubs.onlinelibrary.wiley.com/doi/abs/10.1111/j.1749-6632.1973.tb41447.x}{Naked
  singularities}, Annals of the New York Academy of Sciences 224~(1) (1973)
  125--134.
\newline\urlprefix\url{https://nyaspubs.onlinelibrary.wiley.com/doi/abs/10.1111/j.1749-6632.1973.tb41447.x}

\bibitem{shapiro1991formation}
S.~L. Shapiro, S.~A. Teukolsky,
  \href{https://journals.aps.org/prl/abstract/10.1103/PhysRevLett.66.994}{Formation
  of naked singularities: the violation of cosmic censorship}, Physical Review
  Letters 66~(8) (1991) 994.
\newline\urlprefix\url{https://journals.aps.org/prl/abstract/10.1103/PhysRevLett.66.994}

\bibitem{joshi1992structure}
P.~Joshi, I.~Dwivedi,
  \href{https://link.springer.com/article/10.1007/BF02102631}{The structure of
  naked singularity in self-similar gravitational collapse}, Communications in
  mathematical physics 146 (1992) 333--342.
\newblock \href {http://arxiv.org/abs/gr-qc/9302008}
  {\path{arXiv:gr-qc/9302008}}.
\newline\urlprefix\url{https://link.springer.com/article/10.1007/BF02102631}

\bibitem{christodoulou1994examples}
D.~Christodoulou, \href{https://www.jstor.org/stable/2118619}{Examples of naked
  singularity formation in the gravitational collapse of a scalar field},
  Annals of Mathematics 140~(3) (1994) 607--653.
\newline\urlprefix\url{https://www.jstor.org/stable/2118619}

\bibitem{de2001turning}
F.~de~Felice, Y.~Yunqiang,
  \href{https://dx.doi.org/10.1088/0264-9381/18/7/307}{Turning a black hole
  into a naked singularity}, Classical and Quantum Gravity 18~(7) (2001) 1235.
\newline\urlprefix\url{https://dx.doi.org/10.1088/0264-9381/18/7/307}

\bibitem{harada2002physical}
T.~Harada, H.~Iguchi, K.-i. Nakao,
  \href{https://academic.oup.com/ptp/article-abstract/107/3/449/1809530}{Physical
  processes in naked singularity formation}, Progress of Theoretical Physics
  107~(3) (2002) 449--524.
\newblock \href {http://arxiv.org/abs/gr-qc/0204008}
  {\path{arXiv:gr-qc/0204008}}.
\newline\urlprefix\url{https://academic.oup.com/ptp/article-abstract/107/3/449/1809530}

\bibitem{goswami2006quantum}
R.~Goswami, P.~S. Joshi, P.~Singh,
  \href{https://journals.aps.org/prl/abstract/10.1103/PhysRevLett.96.031302}{Quantum
  evaporation of a naked singularity}, Physical Review Letters 96~(3) (2006)
  031302.
\newblock \href {http://arxiv.org/abs/gr-qc/0506129}
  {\path{arXiv:gr-qc/0506129}}.
\newline\urlprefix\url{https://journals.aps.org/prl/abstract/10.1103/PhysRevLett.96.031302}

\bibitem{joshi2009naked}
P.~S. Joshi, \href{https://www.jstor.org/stable/26001219}{Naked singularities},
  Scientific American 300~(2) (2009) 36--43.
\newline\urlprefix\url{https://www.jstor.org/stable/26001219}

\bibitem{hawking1971gravitationally}
S.~Hawking,
  \href{https://academic.oup.com/mnras/article/152/1/75/2604549}{Gravitationally
  collapsed objects of very low mass}, Monthly Notices of the Royal
  Astronomical Society 152~(1) (1971) 75--78.
\newline\urlprefix\url{https://academic.oup.com/mnras/article/152/1/75/2604549}

\bibitem{sorkin2001formation}
E.~Sorkin, T.~Piran,
  \href{https://journals.aps.org/prd/abstract/10.1103/PhysRevD.63.124024}{Formation
  and evaporation of charged black holes}, Physical Review D 63~(12) (2001)
  124024.
\newblock \href {http://arxiv.org/abs/gr-qc/0103090}
  {\path{arXiv:gr-qc/0103090}}.
\newline\urlprefix\url{https://journals.aps.org/prd/abstract/10.1103/PhysRevD.63.124024}

\bibitem{ray2003electrically}
S.~Ray, A.~L. Espindola, M.~Malheiro, J.~P. Lemos, V.~T. Zanchin,
  \href{https://journals.aps.org/prd/abstract/10.1103/PhysRevD.68.084004}{Electrically
  charged compact stars and formation of charged black holes}, Physical Review
  D 68~(8) (2003) 084004.
\newblock \href {http://arxiv.org/abs/astro-ph/0307262}
  {\path{arXiv:astro-ph/0307262}}.
\newline\urlprefix\url{https://journals.aps.org/prd/abstract/10.1103/PhysRevD.68.084004}

\bibitem{cuesta2003charge}
H.~J.~M. Cuesta, A.~Penna-Firme, A.~P{\'e}rez-Lorenzana,
  \href{https://journals.aps.org/prd/abstract/10.1103/PhysRevD.67.087702}{Charge
  asymmetry in the brane world and the formation of charged black holes},
  Physical Review D 67~(8) (2003) 087702.
\newblock \href {http://arxiv.org/abs/hep-ph/0203010}
  {\path{arXiv:hep-ph/0203010}}.
\newline\urlprefix\url{https://journals.aps.org/prd/abstract/10.1103/PhysRevD.67.087702}

\bibitem{hwang2011internal}
D.-I. Hwang, D.-H. Yeom,
  \href{https://journals.aps.org/prd/abstract/10.1103/PhysRevD.84.064020}{Internal
  structure of charged black holes}, Physical Review D 84~(6) (2011) 064020.
\newblock \href {http://arxiv.org/abs/1010.2585} {\path{arXiv:1010.2585}}.
\newline\urlprefix\url{https://journals.aps.org/prd/abstract/10.1103/PhysRevD.84.064020}

\bibitem{hwang2012dynamical}
D.-I. Hwang, H.~Kim, D.-H. Yeom,
  \href{https://iopscience.iop.org/article/10.1088/0264-9381/29/5/055003/meta}{Dynamical
  formation and evolution of (2+ 1)-dimensional charged black holes}, Classical
  and Quantum Gravity 29~(5) (2012) 055003.
\newblock \href {http://arxiv.org/abs/1105.1371} {\path{arXiv:1105.1371}}.
\newline\urlprefix\url{https://iopscience.iop.org/article/10.1088/0264-9381/29/5/055003/meta}

\bibitem{newman1963empty}
E.~Newman, L.~Tamburino, T.~Unti,
  \href{https://pubs.aip.org/aip/jmp/article-abstract/4/7/915/230250/Empty-Space-Generalization-of-the-Schwarzschild}{Empty-space
  generalization of the schwarzschild metric}, Journal of Mathematical Physics
  4~(7) (1963) 915--923.
\newline\urlprefix\url{https://pubs.aip.org/aip/jmp/article-abstract/4/7/915/230250/Empty-Space-Generalization-of-the-Schwarzschild}

\bibitem{kramer1983exact}
D.~Kramer, H.~Stephani,
  \href{https://ui.adsabs.harvard.edu/abs/1983grg..conf...75K/abstract}{Exact
  solutions of {Einstein's} field equations}, General Relativity and
  Gravitation 1980 (1983) 75.
\newline\urlprefix\url{https://ui.adsabs.harvard.edu/abs/1983grg..conf...75K/abstract}

\bibitem{kerr1963gravitational}
R.~P. Kerr, \href{https://doi.org/10.1103/PhysRevLett.11.237}{Gravitational
  field of a spinning mass as an example of algebraically special metrics},
  Physical review letters 11~(5) (1963) 237.
\newline\urlprefix\url{https://doi.org/10.1103/PhysRevLett.11.237}

\bibitem{newman2014kerr}
E.~T. Newman, T.~Adamo, \href{doi:10.4249/scholarpedia.31791}{Kerr-{Newman}
  metric}, Scholarpedia 9~(10) (2014) 31791.
\newline\urlprefix\url{doi:10.4249/scholarpedia.31791}

\bibitem{ramaswamy1986comment}
S.~Ramaswamy, A.~Sen,
  \href{https://doi.org/10.1103/PhysRevLett.57.1088}{Comment on"
  gravitomagnetic pole and mass quantization"}, Physical Review Letters 57~(8)
  (1986) 1088.
\newline\urlprefix\url{https://doi.org/10.1103/PhysRevLett.57.1088}

\bibitem{misner1963flatter}
C.~W. Misner,
  \href{https://pubs.aip.org/aip/jmp/article-abstract/4/7/924/230241}{The
  flatter regions of {Newman, Unti, and Tamburino's} generalized
  {Schwarzschild} space}, Journal of Mathematical Physics 4~(7) (1963)
  924--937.
\newline\urlprefix\url{https://pubs.aip.org/aip/jmp/article-abstract/4/7/924/230241}

\bibitem{misner1967contribution}
C.~Misner, \href{https://bookstore.ams.org/lam}{Contribution to lectures in
  applied mathematics, vol. 8}, in: Am. Math. Soc, 1967, p. 160.
\newline\urlprefix\url{https://bookstore.ams.org/lam}

\bibitem{bonnor1969new}
W.~B. Bonnor,
  \href{https://www.cambridge.org/core/journals/mathematical-proceedings-of-the-cambridge-philosophical-society/article/new-interpretation-of-the-nut-metric-in-general-relativity/3C7F5F3E0DC2F1B355F2B0D195EEE0D1}{A
  new interpretation of the {NUT} metric in general relativity}, in:
  Mathematical Proceedings of the Cambridge Philosophical Society, Vol.~66,
  Cambridge University Press, 1969, pp. 145--151.
\newline\urlprefix\url{https://www.cambridge.org/core/journals/mathematical-proceedings-of-the-cambridge-philosophical-society/article/new-interpretation-of-the-nut-metric-in-general-relativity/3C7F5F3E0DC2F1B355F2B0D195EEE0D1}

\bibitem{manko2005physical}
V.~Manko, E.~Ruiz,
  \href{https://iopscience.iop.org/article/10.1088/0264-9381/22/17/014/meta}{Physical
  interpretation of the {NUT} family of solutions}, Classical and Quantum
  Gravity 22~(17) (2005) 3555.
\newline\urlprefix\url{https://iopscience.iop.org/article/10.1088/0264-9381/22/17/014/meta}

\bibitem{zimmerman1989geodesics}
R.~L. Zimmerman, B.~Y. Shahir,
  \href{https://doi.org/10.1007/BF00758986}{Geodesics for the {NUT} metric and
  gravitational monopoles}, General relativity and gravitation 21 (1989)
  821--848.
\newline\urlprefix\url{https://doi.org/10.1007/BF00758986}

\bibitem{lynden1998classical}
D.~Lynden-Bell, M.~Nouri-Zonoz,
  \href{https://doi.org/10.1103/RevModPhys.70.427}{Classical monopoles:
  {Newton, NUT} space, gravomagnetic lensing, and atomic spectra}, Reviews of
  Modern Physics 70~(2) (1998) 427.
\newblock \href {http://arxiv.org/abs/gr-qc/9612049}
  {\path{arXiv:gr-qc/9612049}}.
\newline\urlprefix\url{https://doi.org/10.1103/RevModPhys.70.427}

\bibitem{nouri1997gravomagnetic}
M.~Nouri-Zonoz, D.~Lynden-Bell,
  \href{https://doi.org/10.1093/mnras/292.3.714}{Gravomagnetic lensing by {NUT}
  space}, Monthly Notices of the Royal Astronomical Society 292~(3) (1997)
  714--722.
\newline\urlprefix\url{https://doi.org/10.1093/mnras/292.3.714}

\bibitem{rahvar2003gravitational}
S.~Rahvar, M.~Nouri-Zonoz,
  \href{https://doi.org/10.1046/j.1365-8711.2003.06137.x}{Gravitational
  microlensing in {NUT} space}, Monthly Notices of the Royal Astronomical
  Society 338~(4) (2003) 926--930.
\newblock \href {http://arxiv.org/abs/astro-ph/0204282}
  {\path{arXiv:astro-ph/0204282}}.
\newline\urlprefix\url{https://doi.org/10.1046/j.1365-8711.2003.06137.x}

\bibitem{chakraborty2018does}
C.~Chakraborty, S.~Bhattacharyya,
  \href{https://doi.org/10.1103/PhysRevD.98.043021}{Does the gravitomagnetic
  monopole exist? a clue from a black hole x-ray binary}, Physical Review D
  98~(4) (2018) 043021.
\newblock \href {http://arxiv.org/abs/1712.01156} {\path{arXiv:1712.01156}}.
\newline\urlprefix\url{https://doi.org/10.1103/PhysRevD.98.043021}

\bibitem{hawking1977gravitational}
S.~W. Hawking,
  \href{https://doi.org/10.1016/0375-9601(77)90386-3}{Gravitational
  instantons}, Physics Letters A 60~(2) (1977) 81--83.
\newline\urlprefix\url{https://doi.org/10.1016/0375-9601(77)90386-3}

\bibitem{arratia2021hairy}
E.~Arratia, C.~Corral, J.~Figueroa, L.~Sanhueza,
  \href{https://doi.org/10.1103/PhysRevD.103.064068}{{Hairy
  Taub-NUT}/bolt-{AdS} solutions in {Horndeski} theory}, Physical Review D
  103~(6) (2021) 064068.
\newblock \href {http://arxiv.org/abs/2010.02460} {\path{arXiv:2010.02460}}.
\newline\urlprefix\url{https://doi.org/10.1103/PhysRevD.103.064068}

\bibitem{barrientos2022gravitational}
J.~Barrientos, A.~Cisterna, C.~Corral, M.~Oyarzo,
  \href{https://doi.org/10.1007/JHEP05(2022)110}{Gravitational instantons with
  conformally coupled scalar fields}, Journal of High Energy Physics 2022~(5)
  (2022) 1--28.
\newblock \href {http://arxiv.org/abs/2202.13854} {\path{arXiv:2202.13854}}.
\newline\urlprefix\url{https://doi.org/10.1007/JHEP05(2022)110}

\bibitem{brill1964electromagnetic}
D.~R. Brill, \href{https://doi.org/10.1103/PhysRev.133.B845}{Electromagnetic
  fields in a homogeneous, nonisotropic universe}, Physical Review 133~(3B)
  (1964) B845.
\newline\urlprefix\url{https://doi.org/10.1103/PhysRev.133.B845}

\bibitem{ernst1968new}
F.~J. Ernst,
  \href{https://journals.aps.org/pr/abstract/10.1103/PhysRev.167.1175}{New
  formulation of the axially symmetric gravitational field problem}, Physical
  Review 167~(5) (1968) 1175.
\newline\urlprefix\url{https://journals.aps.org/pr/abstract/10.1103/PhysRev.167.1175}

\bibitem{ernst1968new2}
F.~J. Ernst,
  \href{https://journals.aps.org/pr/abstract/10.1103/PhysRev.168.1415}{New
  formulation of the axially symmetric gravitational field problem. ii},
  Physical Review 168~(5) (1968) 1415.
\newline\urlprefix\url{https://journals.aps.org/pr/abstract/10.1103/PhysRev.168.1415}

\bibitem{astorino2013embedding}
M.~Astorino, \href{https://doi.org/10.1103/PhysRevD.87.084029}{Embedding hairy
  black holes in a magnetic universe}, Physical Review D 87~(8) (2013) 084029.
\newblock \href {http://arxiv.org/abs/1301.6794} {\path{arXiv:1301.6794}}.
\newline\urlprefix\url{https://doi.org/10.1103/PhysRevD.87.084029}

\bibitem{astorino2015stationary}
M.~Astorino, \href{https://doi.org/10.1103/PhysRevD.91.064066}{Stationary
  axisymmetric spacetimes with a conformally coupled scalar field}, Physical
  Review D 91~(6) (2015) 064066.
\newblock \href {http://arxiv.org/abs/1412.3539} {\path{arXiv:1412.3539}}.
\newline\urlprefix\url{https://doi.org/10.1103/PhysRevD.91.064066}

\bibitem{barrientos2023ehlers}
J.~Barrientos, A.~Cisterna,
  \href{https://doi.org/10.1103/PhysRevD.108.024059}{Ehlers transformations as
  a tool for constructing accelerating nut black holes}, Physical Review D
  108~(2) (2023) 024059.
\newblock \href {http://arxiv.org/abs/2305.03765} {\path{arXiv:2305.03765}}.
\newline\urlprefix\url{https://doi.org/10.1103/PhysRevD.108.024059}

\bibitem{cisterna2023exact}
A.~Cisterna, K.~M{\"u}ller, K.~Pallikaris, A.~Vigan{\`o},
  \href{https://doi.org/10.1103/PhysRevD.108.024066}{Exact rotating wormholes
  via {Ehlers} transformations}, Physical Review D 108~(2) (2023) 024066.
\newblock \href {http://arxiv.org/abs/2306.14541} {\path{arXiv:2306.14541}}.
\newline\urlprefix\url{https://doi.org/10.1103/PhysRevD.108.024066}

\bibitem{astorino2020enhanced}
M.~Astorino, \href{https://doi.org/10.1007/JHEP01(2020)123}{Enhanced {Ehlers
  transformation and the Majumdar-Papapetrou-NUT} spacetime}, Journal of High
  Energy Physics 2020~(1) (2020) 1--38.
\newblock \href {http://arxiv.org/abs/1906.08228} {\path{arXiv:1906.08228}}.
\newline\urlprefix\url{https://doi.org/10.1007/JHEP01(2020)123}

\bibitem{darmois1927memorial}
G.~Darmois,
  \href{http://www.numdam.org/series/%22M%C3%A9morial%20des%20sciences%20math%C3%A9matiques%22-p/}{M{\'e}morial
  des sciences math{\'e}matiques}, Fascicule XXV (Gauthier-Villars, Paris,
  1927) (1927).
\newline\urlprefix\url{http://www.numdam.org/series/%22M%C3%A9morial%20des%20sciences%20math%C3%A9matiques%22-p/}

\bibitem{erez1959gravitational}
G.~Erez, N.~Rosen, \href{https://www.osti.gov/biblio/4201189}{The gravitational
  field of a particle possessing a multipole moment}, Tech. rep., Israel Inst.
  of Tech., Haifa (1959).
\newline\urlprefix\url{https://www.osti.gov/biblio/4201189}

\bibitem{zipoy1966topology}
D.~M. Zipoy,
  \href{https://pubs.aip.org/aip/jmp/article/7/6/1137/390889/Topology-of-Some-Spheroidal-Metrics}{Topology
  of some spheroidal metrics}, Journal of Mathematical Physics 7~(6) (1966)
  1137--1143.
\newline\urlprefix\url{https://pubs.aip.org/aip/jmp/article/7/6/1137/390889/Topology-of-Some-Spheroidal-Metrics}

\bibitem{voorhees1970static}
B.~Voorhees,
  \href{https://journals.aps.org/prd/abstract/10.1103/PhysRevD.2.2119}{Static
  axially symmetric gravitational fields}, Physical Review D 2~(10) (1970)
  2119.
\newline\urlprefix\url{https://journals.aps.org/prd/abstract/10.1103/PhysRevD.2.2119}

\bibitem{lora2023q}
F.~Lora-Clavijo, G.~Prada-M{\'e}ndez, L.~Becerra, E.~Becerra-Vergara,
  \href{10.1088/1361-6382/ad0b9e}{The q-metric naked singularity: a viable
  explanation for the nature of the central object in the milky way}, Classical
  and Quantum Gravity 40~(24) (2023) 245012.
\newblock \href {http://arxiv.org/abs/2311.06653} {\path{arXiv:2311.06653}}.
\newline\urlprefix\url{10.1088/1361-6382/ad0b9e}

\bibitem{destounis2023geodesics}
K.~Destounis, G.~Huez, K.~D. Kokkotas,
  \href{https://doi.org/10.1007/s10714-023-03119-2}{Geodesics and gravitational
  waves in chaotic extreme-mass-ratio inspirals: the curious case of
  {Zipoy-Voorhees} black-hole mimickers}, General Relativity and Gravitation
  55~(6) (2023) 71.
\newblock \href {http://arxiv.org/abs/2301.11483} {\path{arXiv:2301.11483}}.
\newline\urlprefix\url{https://doi.org/10.1007/s10714-023-03119-2}

\bibitem{richterek2002einstein}
L.~Richterek, J.~Novotn{\`y}, J.~Horsk{\`y},
  \href{https://doi.org/10.1023/A:1020581415399}{{Einstein-Maxwell} fields
  generated from the $\gamma$-metric and their limits}, Czechoslovak journal of
  physics 52 (2002) 1021--1040.
\newblock \href {http://arxiv.org/abs/gr-qc/0209094}
  {\path{arXiv:gr-qc/0209094}}.
\newline\urlprefix\url{https://doi.org/10.1023/A:1020581415399}

\bibitem{chakrabarty2018unattainable}
H.~Chakrabarty, C.~A. Benavides-Gallego, C.~Bambi, L.~Modesto,
  \href{https://doi.org/10.1007/JHEP03(2018)013}{Unattainable extended
  spacetime regions in conformal gravity}, Journal of High Energy Physics
  2018~(3) (2018) 1--12.
\newblock \href {http://arxiv.org/abs/1711.07198} {\path{arXiv:1711.07198}}.
\newline\urlprefix\url{https://doi.org/10.1007/JHEP03(2018)013}

\bibitem{toshmatov2019harmonic}
B.~Toshmatov, D.~Malafarina, N.~Dadhich,
  \href{https://doi.org/10.1103/PhysRevD.100.044001}{Harmonic oscillations of
  neutral particles in the $\gamma$ metric}, Physical Review D 100~(4) (2019)
  044001.
\newblock \href {http://arxiv.org/abs/1905.01088} {\path{arXiv:1905.01088}}.
\newline\urlprefix\url{https://doi.org/10.1103/PhysRevD.100.044001}

\bibitem{allahyari2019quasinormal}
A.~Allahyari, H.~Firouzjahi, B.~Mashhoon,
  \href{https://doi.org/10.1103/PhysRevD.99.044005}{Quasinormal modes of a
  black hole with quadrupole moment}, Physical Review D 99~(4) (2019) 044005.
\newblock \href {http://arxiv.org/abs/1812.03376} {\path{arXiv:1812.03376}}.
\newline\urlprefix\url{https://doi.org/10.1103/PhysRevD.99.044005}

\bibitem{chakrabarty2022effects}
H.~Chakrabarty, D.~Borah, A.~Abdujabbarov, D.~Malafarina, B.~Ahmedov,
  \href{https://doi.org/10.1140/epjc/s10052-021-09982-0}{Effects of
  gravitational lensing on neutrino oscillation in $\gamma$-spacetime}, The
  European Physical Journal C 82~(1) (2022) 24.
\newline\urlprefix\url{https://doi.org/10.1140/epjc/s10052-021-09982-0}

\bibitem{fisher1999scalar}
I.~Fisher, \href{https://arxiv.org/abs/gr-qc/9911008}{Scalar mesostatic field
  with regard for gravitational effects}, Zh. Eksp. Teor. Fiz 18 (1948) 636.
\newblock \href {http://arxiv.org/abs/gr-qc/9911008}
  {\path{arXiv:gr-qc/9911008}}.
\newline\urlprefix\url{https://arxiv.org/abs/gr-qc/9911008}

\bibitem{janis1968reality}
A.~I. Janis, E.~T. Newman, J.~Winicour,
  \href{https://journals.aps.org/prl/abstract/10.1103/PhysRevLett.20.878}{Reality
  of the {Schwarzschild} singularity}, Physical Review Letters 20~(16) (1968)
  878.
\newline\urlprefix\url{https://journals.aps.org/prl/abstract/10.1103/PhysRevLett.20.878}

\bibitem{wyman1981static}
M.~Wyman,
  \href{https://journals.aps.org/prd/abstract/10.1103/PhysRevD.24.839}{Static
  spherically symmetric scalar fields in general relativity}, Physical Review D
  24~(4) (1981) 839.
\newline\urlprefix\url{https://journals.aps.org/prd/abstract/10.1103/PhysRevD.24.839}

\bibitem{virbhadra2002gravitational}
K.~S. Virbhadra, G.~F. Ellis,
  \href{https://doi.org/10.1103/PhysRevD.65.103004}{Gravitational lensing by
  naked singularities}, Physical Review D 65~(10) (2002) 103004.
\newline\urlprefix\url{https://doi.org/10.1103/PhysRevD.65.103004}

\bibitem{chen2024gravitational}
D.~Chen, Y.~Chen, P.~Wang, T.~Wu, H.~Wu,
  \href{https://doi.org/10.1140/epjc/s10052-024-12950-z}{Gravitational lensing
  by transparent {Janis--Newman--Winicour} naked singularities}, The European
  Physical Journal C 84~(6) (2024) 584.
\newblock \href {http://arxiv.org/abs/2309.00905} {\path{arXiv:2309.00905}}.
\newline\urlprefix\url{https://doi.org/10.1140/epjc/s10052-024-12950-z}

\bibitem{solanki2022shadows}
D.~N. Solanki, P.~Bambhaniya, D.~Dey, P.~S. Joshi, K.~N. Pathak,
  \href{https://doi.org/10.1140/epjc/s10052-022-10045-1}{Shadows and precession
  of orbits in rotating {Janis--Newman--Winicour} spacetime}, The European
  Physical Journal C 82~(1) (2022) 77.
\newblock \href {http://arxiv.org/abs/2109.14937} {\path{arXiv:2109.14937}}.
\newline\urlprefix\url{https://doi.org/10.1140/epjc/s10052-022-10045-1}

\bibitem{stashko2024quasinormal}
O.~Stashko, O.~Savchuk, V.~Zhdanov,
  \href{https://doi.org/10.1103/PhysRevD.109.024012}{Quasinormal modes of naked
  singularities in presence of nonlinear scalar fields}, Physical Review D
  109~(2) (2024) 024012.
\newblock \href {http://arxiv.org/abs/2307.04295} {\path{arXiv:2307.04295}}.
\newline\urlprefix\url{https://doi.org/10.1103/PhysRevD.109.024012}

\bibitem{virbhadra1997nature}
K.~Virbhadra, S.~Jhingan, P.~Joshi,
  \href{https://doi.org/10.1142/S0218271897000200}{Nature of singularity in
  {Einstein}-massless scalar theory}, International Journal of Modern Physics D
  6~(03) (1997) 357--361.
\newblock \href {http://arxiv.org/abs/gr-qc/9512030}
  {\path{arXiv:gr-qc/9512030}}.
\newline\urlprefix\url{https://doi.org/10.1142/S0218271897000200}

\bibitem{chowdhury2011circular}
A.~N. Chowdhury, M.~Patil, D.~Malafarina, P.~S. Joshi,
  \href{https://doi.org/10.1103/PhysRevD.85.104031}{Circular geodesics and
  accretion disks in {Janis-Newman-Winicour} and gamma metric}, Physical Review
  D (2011).
\newblock \href {http://arxiv.org/abs/1112.2522} {\path{arXiv:1112.2522}}.
\newline\urlprefix\url{https://doi.org/10.1103/PhysRevD.85.104031}

\bibitem{azizallahi2024three}
A.~Azizallahi, B.~Mirza, A.~Hajibarat, H.~Anjomshoa,
  \href{https://www.sciencedirect.com/science/article/pii/S0550321323003413}{Three
  parameter metrics in the presence of a scalar field in four and higher
  dimensions}, Nuclear Physics B 998 (2024) 116414.
\newblock \href {http://arxiv.org/abs/arXiv:2307.09328}
  {\path{arXiv:arXiv:2307.09328}}.
\newline\urlprefix\url{https://www.sciencedirect.com/science/article/pii/S0550321323003413}

\bibitem{mirza2023class}
B.~Mirza, P.~K. Kangazi, F.~Sadeghi,
  \href{https://link.springer.com/article/10.1140/epjc/s10052-023-12255-7}{A
  class of rotating metrics in the presence of a scalar field}, The European
  Physical Journal C 83~(12) (2023) 1--12.
\newblock \href {http://arxiv.org/abs/2307.13588} {\path{arXiv:2307.13588}}.
\newline\urlprefix\url{https://link.springer.com/article/10.1140/epjc/s10052-023-12255-7}

\bibitem{toktarbay2014stationary}
S.~Toktarbay, H.~Quevedo,
  \href{https://link.springer.com/article/10.1134/S0202289314040136}{A
  stationary q-metric}, Gravit. Cosmol. 20~(4) (2014) 252--254.
\newblock \href {http://arxiv.org/abs/1510.04155} {\path{arXiv:1510.04155}}.
\newline\urlprefix\url{https://link.springer.com/article/10.1134/S0202289314040136}

\bibitem{frutos2018relativistic}
F.~Frutos-Alfaro, M.~Soffel,
  \href{https://royalsocietypublishing.org/doi/full/10.1098/rsos.180640}{On
  relativistic multipole moments of stationary space--times}, R. Soc. Open Sci.
  5~(7) (2018) 180640.
\newblock \href {http://arxiv.org/abs/1606.07173} {\path{arXiv:1606.07173}}.
\newline\urlprefix\url{https://royalsocietypublishing.org/doi/full/10.1098/rsos.180640}

\bibitem{bogush2020generation}
I.~Bogush, D.~Gal’tsov,
  \href{https://journals.aps.org/prd/abstract/10.1103/PhysRevD.102.124006}{Generation
  of rotating solutions in {Einstein}-scalar gravity}, Phys. Rev. D 102~(12)
  (2020) 124006.
\newblock \href {http://arxiv.org/abs/2001.02936} {\path{arXiv:2001.02936}}.
\newline\urlprefix\url{https://journals.aps.org/prd/abstract/10.1103/PhysRevD.102.124006}

\bibitem{mazharimousavi2023nonspherically}
S.~H. Mazharimousavi,
  \href{https://doi.org/10.1140/epjc/s10052-023-12300-5}{Nonspherically-symmetric
  black hole in {Einstein}-massless scalar theory}, The European Physical
  Journal C 83 (2023) 1131.
\newblock \href {http://arxiv.org/abs/2309.06571} {\path{arXiv:2309.06571}}.
\newline\urlprefix\url{https://doi.org/10.1140/epjc/s10052-023-12300-5}

\bibitem{harrison1968new}
B.~K. Harrison,
  \href{https://pubs.aip.org/aip/jmp/article-abstract/9/11/1744/1031163}{New
  solutions of the {Einstein}-{Maxwell} equations from old}, Journal of
  Mathematical Physics 9~(11) (1968) 1744--1752.
\newline\urlprefix\url{https://pubs.aip.org/aip/jmp/article-abstract/9/11/1744/1031163}

\bibitem{ehlers1958konstruktionen}
J.~Ehlers,
  \href{https://pure.mpg.de/pubman/faces/ViewItemOverviewPage.jsp?itemId=item_153700}{Konstruktionen
  und charakterisierungen von l{\"o}sungen der {Einsteinschen}
  gravitationsfeldgleichungen}, Ph.D. thesis, Hamburg Hamburg, Germany (1958).
\newline\urlprefix\url{https://pure.mpg.de/pubman/faces/ViewItemOverviewPage.jsp?itemId=item_153700}

\bibitem{schneider1992gravitational}
P.~Schneider, J.~Ehlers, E.~E. Falco, P.~Schneider, J.~Ehlers, E.~E. Falco,
  \href{https://link.springer.com/chapter/10.1007/978-3-662-03758-4_13}{Gravitational
  lenses as astrophysical tools}, Gravitational Lenses (1992) 467--515.
\newline\urlprefix\url{https://link.springer.com/chapter/10.1007/978-3-662-03758-4_13}

\bibitem{wambsganss1998gravitational}
J.~Wambsganss,
  \href{https://link.springer.com/article/10.12942/lrr-1998-12}{Gravitational
  lensing in astronomy}, Living Reviews in Relativity 1 (1998) 1--74.
\newblock \href {http://arxiv.org/abs/astro-ph/9812021}
  {\path{arXiv:astro-ph/9812021}}.
\newline\urlprefix\url{https://link.springer.com/article/10.12942/lrr-1998-12}

\bibitem{narayan1999gravitational}
R.~Narayan, M.~Bartelmann,
  \href{https://ui.adsabs.harvard.edu/abs/1999fsu..conf..360N/abstract}{Gravitational
  lensing}, Formation of Structure in the Universe (1999) 360\href
  {http://arxiv.org/abs/1010.3829} {\path{arXiv:1010.3829}}.
\newline\urlprefix\url{https://ui.adsabs.harvard.edu/abs/1999fsu..conf..360N/abstract}

\bibitem{schneider2006gravitational}
P.~Schneider, C.~Kochanek, J.~Wambsganss,
  \href{https://books.google.com/books?hl=en&lr=&id=AF8-ErlCb94C&oi=fnd&pg=PA1&dq=Schneider+P,+Kochanek+C+S+and+Wambsganss+J+2006+Gravitational+Lensing:+Strong,+Weak+and+Micro+(Berlin:+Springer)&ots=1zJ6JvvTIl&sig=o4brS5R3nADak1UJWg_Vg7U_uN0#v=onepage&q=Schneider%20P%2C%20Kochanek%20C%20S%20and%20Wambsganss%20J%202006%20Gravitational%20Lensing%3A%20Strong%2C%20Weak%20and%20Micro%20(Berlin%3A%20Springer)&f=false}{Gravitational
  lensing: strong, weak and micro: Saas-Fee advanced course 33}, Vol.~33,
  Springer Science \& Business Media, 2006.
\newline\urlprefix\url{https://books.google.com/books?hl=en&lr=&id=AF8-ErlCb94C&oi=fnd&pg=PA1&dq=Schneider+P,+Kochanek+C+S+and+Wambsganss+J+2006+Gravitational+Lensing:+Strong,+Weak+and+Micro+(Berlin:+Springer)&ots=1zJ6JvvTIl&sig=o4brS5R3nADak1UJWg_Vg7U_uN0#v=onepage&q=Schneider%20P%2C%20Kochanek%20C%20S%20and%20Wambsganss%20J%202006%20Gravitational%20Lensing%3A%20Strong%2C%20Weak%20and%20Micro%20(Berlin%3A%20Springer)&f=false}

\bibitem{straumann2012general}
N.~Straumann,
  \href{https://books.google.com/books?hl=en&lr=&id=jjBMw0KFtZgC&oi=fnd&pg=PR3&dq=Norbert+Straumann+General+relativity&ots=E7An20VgBm&sig=q8ksfy1Q83lafVV6YBrXoY3ydOk#v=onepage&q=Norbert%20Straumann%20General%20relativity&f=false}{General
  relativity}, Springer Science \& Business Media, 2012.
\newline\urlprefix\url{https://books.google.com/books?hl=en&lr=&id=jjBMw0KFtZgC&oi=fnd&pg=PR3&dq=Norbert+Straumann+General+relativity&ots=E7An20VgBm&sig=q8ksfy1Q83lafVV6YBrXoY3ydOk#v=onepage&q=Norbert%20Straumann%20General%20relativity&f=false}

\bibitem{nollert1996significance}
H.-P. Nollert,
  \href{https://journals.aps.org/prd/abstract/10.1103/PhysRevD.53.4397}{About
  the significance of quasinormal modes of black holes}, Physical Review D
  53~(8) (1996) 4397.
\newblock \href {http://arxiv.org/abs/gr-qc/9602032}
  {\path{arXiv:gr-qc/9602032}}.
\newline\urlprefix\url{https://journals.aps.org/prd/abstract/10.1103/PhysRevD.53.4397}

\bibitem{nollert1999quasinormal}
H.-P. Nollert,
  \href{https://iopscience.iop.org/article/10.1088/0264-9381/16/12/201/meta}{Quasinormal
  modes: the characteristicsound'of black holes and neutron stars}, Classical
  and Quantum Gravity 16~(12) (1999) R159.
\newline\urlprefix\url{https://iopscience.iop.org/article/10.1088/0264-9381/16/12/201/meta}

\bibitem{kokkotas1999quasi}
K.~D. Kokkotas, B.~G. Schmidt,
  \href{https://link.springer.com/article/10.12942/lrr-1999-2}{Quasi-normal
  modes of stars and black holes}, Living Reviews in Relativity 2 (1999) 1--72.
\newblock \href {http://arxiv.org/abs/gr-qc/9909058}
  {\path{arXiv:gr-qc/9909058}}.
\newline\urlprefix\url{https://link.springer.com/article/10.12942/lrr-1999-2}

\bibitem{dreyer2003quasinormal}
O.~Dreyer,
  \href{https://journals.aps.org/prl/abstract/10.1103/PhysRevLett.90.081301}{Quasinormal
  modes, the area spectrum, and black hole entropy}, Physical Review Letters
  90~(8) (2003) 081301.
\newblock \href {http://arxiv.org/abs/gr-qc/0211076}
  {\path{arXiv:gr-qc/0211076}}.
\newline\urlprefix\url{https://journals.aps.org/prl/abstract/10.1103/PhysRevLett.90.081301}

\bibitem{berti2009quasinormal}
E.~Berti, V.~Cardoso, A.~O. Starinets,
  \href{https://iopscience.iop.org/article/10.1088/0264-9381/26/16/163001/meta}{Quasinormal
  modes of black holes and black branes}, Classical and Quantum Gravity 26~(16)
  (2009) 163001.
\newblock \href {http://arxiv.org/abs/0905.2975} {\path{arXiv:0905.2975}}.
\newline\urlprefix\url{https://iopscience.iop.org/article/10.1088/0264-9381/26/16/163001/meta}

\bibitem{denef2010black}
F.~Denef, S.~A. Hartnoll, S.~Sachdev,
  \href{https://iopscience.iop.org/article/10.1088/0264-9381/27/12/125001/meta?casa_token=AIguXZGgOLcAAAAA:NGNAO_S_0lPUWs6yHJSjK9d6ufJnrqctDGbh2Ahp1UXFBbm3vpLbUyV4pdJdd7MPbM5ESHx-BPzk3ZY1YMB-4au9N7UJ}{Black
  hole determinants and quasinormal modes}, Classical and Quantum Gravity
  27~(12) (2010) 125001.
\newblock \href {http://arxiv.org/abs/0908.2657} {\path{arXiv:0908.2657}}.
\newline\urlprefix\url{https://iopscience.iop.org/article/10.1088/0264-9381/27/12/125001/meta?casa_token=AIguXZGgOLcAAAAA:NGNAO_S_0lPUWs6yHJSjK9d6ufJnrqctDGbh2Ahp1UXFBbm3vpLbUyV4pdJdd7MPbM5ESHx-BPzk3ZY1YMB-4au9N7UJ}

\bibitem{konoplya2011quasinormal}
R.~Konoplya, A.~Zhidenko,
  \href{https://journals.aps.org/rmp/abstract/10.1103/RevModPhys.83.793}{Quasinormal
  modes of black holes: From astrophysics to string theory}, Reviews of Modern
  Physics 83~(3) (2011) 793.
\newblock \href {http://arxiv.org/abs/1102.4014} {\path{arXiv:1102.4014}}.
\newline\urlprefix\url{https://journals.aps.org/rmp/abstract/10.1103/RevModPhys.83.793}

\bibitem{flachi2013quasinormal}
A.~Flachi, J.~P. Lemos,
  \href{https://journals.aps.org/prd/abstract/10.1103/PhysRevD.87.024034}{Quasinormal
  modes of regular black holes}, Physical Review D 87~(2) (2013) 024034.
\newblock \href {http://arxiv.org/abs/1211.6212} {\path{arXiv:1211.6212}}.
\newline\urlprefix\url{https://journals.aps.org/prd/abstract/10.1103/PhysRevD.87.024034}

\bibitem{matyjasek2017quasinormal}
J.~Matyjasek, M.~Opala,
  \href{https://journals.aps.org/prd/abstract/10.1103/PhysRevD.96.024011}{Quasinormal
  modes of black holes: The improved semianalytic approach}, Physical Review D
  96~(2) (2017) 024011.
\newblock \href {http://arxiv.org/abs/1704.00361} {\path{arXiv:1704.00361}}.
\newline\urlprefix\url{https://journals.aps.org/prd/abstract/10.1103/PhysRevD.96.024011}

\bibitem{ferrari1984new}
V.~Ferrari, B.~Mashhoon,
  \href{https://journals.aps.org/prd/abstract/10.1103/PhysRevD.30.295}{New
  approach to the quasinormal modes of a black hole}, Physical Review D 30~(2)
  (1984) 295.
\newline\urlprefix\url{https://journals.aps.org/prd/abstract/10.1103/PhysRevD.30.295}

\bibitem{ferrari1984oscillations}
V.~Ferrari, B.~Mashhoon,
  \href{https://journals.aps.org/prl/abstract/10.1103/PhysRevLett.52.1361}{Oscillations
  of a black hole}, Physical Review Letters 52~(16) (1984) 1361.
\newline\urlprefix\url{https://journals.aps.org/prl/abstract/10.1103/PhysRevLett.52.1361}

\bibitem{mashhoon1985stability}
B.~Mashhoon,
  \href{https://journals.aps.org/prd/abstract/10.1103/PhysRevD.31.290}{Stability
  of charged rotating black holes in the eikonal approximation}, Physical
  Review D 31~(2) (1985) 290.
\newline\urlprefix\url{https://journals.aps.org/prd/abstract/10.1103/PhysRevD.31.290}

\bibitem{barrientos2024revisiting}
J.~Barrientos, A.~Cisterna, M.~Hassaine, J.~Oliva,
  \href{https://link.springer.com/article/10.1140/epjc/s10052-024-13383-4}{Revisiting
  {Buchdahl} transformations: new static and rotating black holes in vacuum,
  double copy, and hairy extensions}, The European Physical Journal C 84~(10)
  (2024) 1011.
\newline\urlprefix\url{https://link.springer.com/article/10.1140/epjc/s10052-024-13383-4}

\bibitem{barrientos2024mixing}
J.~Barrientos, A.~Cisterna, I.~Kol{\'a}{\v{r}}, K.~M{\"u}ller, M.~Oyarzo,
  K.~Pallikaris,
  \href{https://link.springer.com/article/10.1140/epjc/s10052-024-13093-x}{Mixing
  {“Magnetic” and “Electric” Ehlers--Harrison} transformations: the
  electromagnetic swirling spacetime and novel type i backgrounds}, The
  European Physical Journal C 84~(7) (2024) 724.
\newline\urlprefix\url{https://link.springer.com/article/10.1140/epjc/s10052-024-13093-x}

\bibitem{gibbons2008applications}
G.~Gibbons, M.~Werner,
  \href{https://dx.doi.org/10.1088/0264-9381/25/23/235009}{Applications of the
  {Gauss--Bonnet} theorem to gravitational lensing}, Classical and Quantum
  Gravity 25~(23) (2008) 235009.
\newblock \href {http://arxiv.org/abs/0807.0854} {\path{arXiv:0807.0854}}.
\newline\urlprefix\url{https://dx.doi.org/10.1088/0264-9381/25/23/235009}

\end{thebibliography}
	\newpage
\end{document}